\documentstyle[psfig]{mn}

\newif\ifAMStwofonts
\AMStwofontstrue

\ifoldfss
  \ifCUPmtlplainloaded \else
    \NewTextAlphabet{textbfit} {cmbxti10} {}
    \NewTextAlphabet{textbfss} {cmssbx10} {}
    \NewMathAlphabet{mathbfit} {cmbxti10} {} 
    \NewMathAlphabet{mathbfss} {cmssbx10} {} 
  \fi
  \ifAMStwofonts
    \ifCUPmtlplainloaded \else
      \NewSymbolFont{upmath} {eurm10}
      \NewSymbolFont{AMSa} {msam10}
      \NewMathSymbol{\upi}     {0}{upmath}{19}
      \NewMathSymbol{\umu}     {0}{upmath}{16}
      \NewMathSymbol{\upartial}{0}{upmath}{40}
      \NewMathSymbol{\leqslant}{3}{AMSa}{36}
      \NewMathSymbol{\geqslant}{3}{AMSa}{3E}

    \fi
  \fi
\fi 

\ifnfssone
  \newmathalphabet{\mathit}
  \addtoversion{normal}{\mathit}{cmr}{m}{it}
  \addtoversion{bold}{\mathit}{cmr}{bx}{it}
  \newmathalphabet{\mathbfit} 
  \addtoversion{normal}{\mathbfit}{cmr}{bx}{it}
  \addtoversion{bold}{\mathbfit}{cmr}{bx}{it}
  \newmathalphabet{\mathbfss} 
  \addtoversion{normal}{\mathbfss}{cmss}{bx}{n}
  \addtoversion{bold}{\mathbfss}{cmss}{bx}{n}
  \ifAMStwofonts
    \ifCUPmtlplainloaded \else
      \UseAMStwoboldmath
      \makeatletter
      \new@mathgroup\upmath@group
      \define@mathgroup\mv@normal\upmath@group{eur}{m}{n}
      \define@mathgroup\mv@bold\upmath@group{eur}{b}{n}
      \edef\UPM{\hexnumber\upmath@group}
      \new@mathgroup\amsa@group
      \define@mathgroup\mv@normal\amsa@group{msa}{m}{n}
      \define@mathgroup\mv@bold\amsa@group{msa}{m}{n}
      \edef\AMSa{\hexnumber\amsa@group}
      \makeatother
      \mathchardef\upi="0\UPM19
      \mathchardef\umu="0\UPM16
      \mathchardef\upartial="0\UPM40
      \mathchardef\leqslant="3\AMSa36
      \mathchardef\geqslant="3\AMSa3E
    \fi
  \fi
\fi 

\ifnfsstwo
  \DeclareMathAlphabet{\mathbfit}{OT1}{cmr}{bx}{it}
  \SetMathAlphabet\mathbfit{bold}{OT1}{cmr}{bx}{it}
  \DeclareMathAlphabet{\mathbfss}{OT1}{cmss}{bx}{n}
  \SetMathAlphabet\mathbfss{bold}{OT1}{cmss}{bx}{n}
  \ifAMStwofonts
    \ifCUPmtlplainloaded \else
      \DeclareSymbolFont{UPM}{U}{eur}{m}{n}
      \SetSymbolFont{UPM}{bold}{U}{eur}{b}{n}
      \DeclareSymbolFont{AMSa}{U}{msa}{m}{n}
      \DeclareMathSymbol{\upi}{0}{UPM}{"19}
      \DeclareMathSymbol{\umu}{0}{UPM}{"16}
      \DeclareMathSymbol{\upartial}{0}{UPM}{"40}
      \DeclareMathSymbol{\leqslant}{3}{AMSa}{"36}
      \DeclareMathSymbol{\geqslant}{3}{AMSa}{"3E}
    \fi
  \fi
\fi 

\ifCUPmtlplainloaded \else
  \ifAMStwofonts \else 
    \def\upi{\pi}
    \def\umu{\mu}
    \def\upartial{\partial}
  \fi
\fi

\title{ASTRO-F - The next generation of mid-infrared surveys}
\author[C.~P~Pearson et al]
       {C.~P.~Pearson$^1$\thanks{further info. contact Chris Pearson (cpp@ir.isas.ac.jp)}, H.~Matsuhara$^1$ T.~Onaka$^2$, H.Watarai$^1$ and T.~Matsumoto$^1$\\
        $^1$Institute of Space and Astronautical Science, Yoshinodai 3-1-1, Sagamihara, Kanagawa 229 8510, Japan\\
        $^2$ Department of Astronomy, School of Science, University of Tokyo, Tokyo 113-0033, Japan}

\date{Accepted .\\
      Received ;\\
      in original form 2000 June 23}
\pagerange{\pageref{firstpage}--\pageref{lastpage}}

\begin{document}

\label{firstpage}

\maketitle

\begin{abstract}

We present basic observational strategies for ASTRO-F (also known as the Imaging Infra Red Surveyor (IRIS)) to be launched in 2004 by the Japanese Institute of Space and Astronautical Science (ISAS). We examine 2 survey scenarios, a deep $\sim$1sq.deg. survey reaching sensitivities an order of magnitude below all but the deepest surveys performed by ISO in the mid-IR, and a shallow $\sim$18sq.deg mid-infrared (7-25$\umu$m in 6 bands) covering an area greater than the entire area covered by all ISO mid-IR surveys. Using 2 cosmological models the number of galaxies predicted for each survey is calculated. The first model uses an enhancement of a {\it classical} $(1+z)^{3.1}$ pure luminosity evolution model ~\cite{cpp96}. The second model incorporates a strongly evolving ULIG component. For the deep survey, between 20,000-30,000 galaxies should be detected in the shortest wavebands and $\approx$5000 in the longest (25$\umu$m) band. It is predicted that the shallow survey will detect of the order of 100,000 - 150,000 sources. We find that for both ASTRO-F and other small aperture space telescopes, confusion due to faint sources may be severe, especially at the longest mid-IR wavelengths. Using the exceptional range of observational options provided by ASTRO-F (9 wavelength filters and spectroscopic ability from 2.2-25$\umu$m), we show that by combining the mid-IR observations with the near-IR camera on ASTRO-F, both the different galaxy populations and rough photometric redshifts can be distinguished in the colour-colour plane. In its role as a surveyor (plus near-IR spectroscopic ability) ASTRO-F will complement well the SIRTF space observatory mission.
\end{abstract}

\begin{keywords}
ASTRO-F/IRIS -- Cosmology: source counts -- Infrared: source counts -- Galaxies: evolution .
\end{keywords}

\section{Introduction}\label{sec:introduction}

Since the launch in 1983 of the extremely successful IRAS mission, infrared astronomy and in particular infrared galaxy surveys, galaxy evolution and cosmology have enjoyed a great deal of exposure and success within the astronomical community. Now rather fittingly, a year after the 20th anniversary of the launch of IRAS the Japanese infrared space telescope ASTRO-F is set to take MID-IR surveys to the next generation. 

Although similar advances in galaxy evolution have been in other wavebands such as at optical wavelengths ~\cite{madau96}, it has become apparent the infrared-sub-mm regime may in fact hold the key to galaxy evolution. The main constraint with optical surveys (HST and ground based) is the extinction due to dust that conceals much of the star formation history of the Universe from optical eyes. At infrared wavelengths the Universe becomes relatively transparent to these extinction and absorption effects providing a unique opportunity to study the star formation history of the Universe relatively unhindered. Furthermore, the recent development of large format infrared arrays combined with a wide field-of-view and high spatial resolution has now been now realized and much better sensitivity than ground-based instruments can be achieved, since observations from space with a cooled telescope do not suffer from any atmospheric and instrument thermal background radiation (especially in the mid-infrared). 

Although the emission from dust due to star formation peaks in the wavelength region between 60-100$\umu$m a significant fraction, about 40 percent, of the luminosity of starburst galaxies is radiated in the mid-infrared region from 8-40$\umu$m ~\cite{soif87}. Furthermore unlike the far-infrared, observations at mid-infrared wavelengths are not so severely constrained by detector array size and resolution.

The IRAS satellite surveyed almost the entire sky at 100, 60, 25 \& 12$\umu$m, detecting more than 25,000 galaxies and providing a huge legacy of data that until the launch of the Infrared Space Observatory ~\cite{kessler96}, remained the only benchmark from which to study galaxy evolution in the infrared. ISO (Nov. 1995 - May 1998) operated at wavelengths from 2.5-240$\umu$m and improved on IRAS by a factor of 1000 in sensitivity and by 100 in angular resolution at 12$\umu$m. Although strictly speaking an observatory, ISO performed numerous galaxy surveys from 200$\umu$m to 7$\umu$m (e.g. Bogun et al. ~\shortcite{bogun96}, Puget et al. ~\shortcite{pug96}, Kawara et al. ~\shortcite{kawara98}, Flores et al. ~\shortcite{flores99a}, ~\shortcite{flores99b}, Taniguchi et al. ~\shortcite{taniguchi97}, Serjeant et al. ~\shortcite{serjeant97}, Oliver et al. ~\shortcite{oliver00a}, Lindern-V{\o}rnle et al. ~\shortcite{vornle00}). 

At 15$\umu$m the ISO surveys covered a wide range in both sensitivity and spatial area. The deepest conventional surveys reaching down to fluxes of between 500-100$\umu$Jy \cite{elbaz99}, although fluctuation analysis in the HDF and the use of clusters as lenses to amplify the source fluxes, extended these sensitivities down almost an order of magnitude further to $\sim$50$\umu$Jy \cite{altieri99}, \cite{metcalfe00}.

Following ISO it was hoped that the Wide Field Infrared Explorer (WIRE) satellite would provide the next generation of mid-infrared surveys, covering 170sq.deg. at 12$\umu$m down to 1.9mJy with smaller area surveys to deeper fluxes. The tragic loss of the WIRE mission left a damaging {\it hole} in infrared astronomy. The Space Infra-Red Telescope Facility (SIRTF) due for launch in 2002 ~\cite{rieke00}, covering the wavelength range from 3.6-160$\umu$m and spectroscopy from 5-37$\umu$m is the next major space infrared telescope although again like ISO before it, SIRTF is officially an observatory. It is hoped that SIRTF and ASTRO-F will complement each other well, both providing their own unique contribution to IR astronomy. 

This paper investigates several basic survey strategies possible in the mid-infrared using the Infra-Red Camera on ASTRO-F. In section ~\ref{sec:ASTRO-F} we introduce the ASTRO-F mission. Section ~\ref{sec:model} explains the model parameters used to predict the survey source counts. In sections ~\ref{sec:survey} and ~\ref{sec:wide} we predict the outcome of a suite of survey options viable with ASTRO-F and in section ~\ref{sec:conclusions} we give a summary of the results.

\begin{figure}
\centering
\centerline{
\psfig{figure=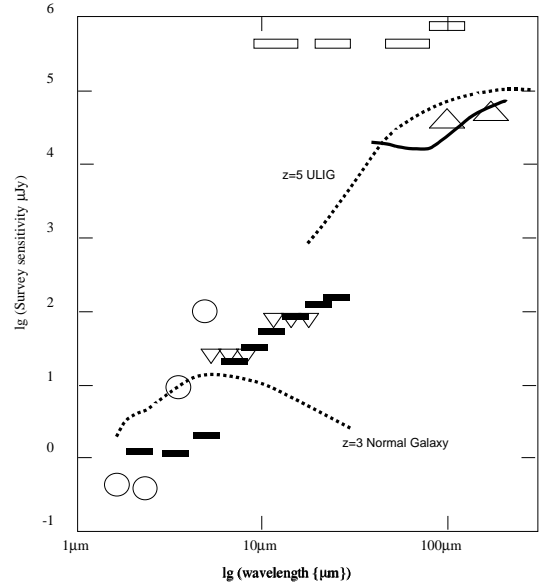,height=10cm}
}
\caption{ASTRO-F survey sensitivities ($5\sigma$) in comparison with other IR detection limits. ASTRO-F FIS and IRC, assuming a 500s integration ({\it solid line and filled boxes\,}), IRAS ({\it empty boxes\,}), ISOPHOT ({\it large upward triangles\,}), ISOCAM Hubble Deep Field survey limits (Goldschmidt 1997) ({\it small downward triangles\,}) and the Japanese Subaru telescope ({\it unfilled circles\,}).
\label{sensitivity}}
\end{figure}

\section{ASTRO-F / The Infra-Red Imaging Surveyor (IRIS)}\label{sec:ASTRO-F}

\subsection{ASTRO-F specifications}

ASTRO-F, also known as the Infra-Red Imaging Surveyor (IRIS, \cite{mura98}) will be the second infrared astronomy mission of the Japanese Institute of Space and Astronautical Science (ISAS). ASTRO-F will be a 70cm cooled telescope and will be dedicated for infrared sky surveys \cite{ona98}. ASTRO-F was preceded by the Infra-Red Telescope in Space, IRTS \cite{mura96}, a short 4 week mission during 1995 optimized to measure the diffuse IR emission/background from 1.5-800$\umu$m \cite{mats00}.

 The ASTRO-F Ritchey-Chertien telescope has a 70cm aperture and is cooled to 6K (the detectors to 1.8K), using a light weight liquid Helium cryostat. The temperature of the outer wall of the cryostat is expected to be maintained below 200K via radiation cooling. Two 2-stage Stirling-cycle coolers ensure minimum heat flow from the outer wall of the cryostat, almost doubling the lifetime of the Helium thus allowing approximately 150l of liquid Helium to sustain the telescope for more than 500 days. Another advantage of using the mechanical coolers is that the near-infrared detectors will still be usable even after the Helium expires (providing the coolers continue to function). The primary and secondary light-weight mirrors (only 11kg) have been made from silicon carbide, adopted due to its large Youngs modulus and high thermal conductivity.

ASTRO-F covers a wide wavelength range from the K-band to 200$\umu$m. Two focal-plane instruments are installed. One is the Far-Infrared Surveyor (\cite{kaw98} \cite{takahashi00}) which will survey the entire sky in the wavelength range from 50 to 200$\umu$m with angular resolutions of 30 - 50 arcsec using high sensitivity Ge:Ga detector arrays (see Takeuchi et al. \shortcite{take99} Pearson et al.\shortcite{cpp004} for FIS survey predictions). The FIS also has a Fourier-transform spectrometer with a resolution of $0.5cm^{-1}$ allowing imaging spectroscopy of selected sources. The other focal-plane instrument is the Infrared Camera (IRC, see Table~\ref{IRC} \cite{mat98},\cite{wat00}). It employs large-format detector arrays and will take deep images of selected sky regions in the near-infrared and mid-infrared range. ASTRO-F will have a much higher sensitivity than that of the IRAS survey. For example, ASTRO-F has 50 - 100 times higher sensitivity at 100$\umu$m and more than 1000 times that at mid-infrared wavelengths. The detection limits are 1-100 mJy in the near-mid infrared and 10-100mJy in the far-infrared (Fig.~\ref{sensitivity}). ASTRO-F will also offer 4 times the resolution over 10 times the detector area compared to ISOCAM on ISO. Although similar sensitivities are also capable with SIRTIF \cite{fazio98}, ASTRO-F will be able to reach such limits over a wider area (10-20sq. deg. due to the larger FOV) in the mid-infrared and the entire sky at far-infrared wavelengths.  At 2.5-5 years, SIRTF has a longer projected lifetime than ASTRO-F, it has no spectroscopic capability at near-infrared wavelengths ($<5\umu m$). ASTRO-F has this added capability and will in fact be able to continue to use this instrument even after liquid Helium exhaustion. 

With the ASTRO-F surveys, great progress is expected in the fields of galaxy evolution, formation of stars and planets, dark matter and brown dwarfs. ASTRO-F is now scheduled to be launched with ISAS's M-V launch vehicle, into a sun-synchronous polar orbit with an altitude of 750 km in mid 2004.

\begin{table*}
\caption{Infra-Red Camera Parameters}
\renewcommand{\arraystretch}{1.4}
\setlength\tabcolsep{15pt}
\begin{tabular}{@{}lp{6.0cm}l}
\hline\noalign{\smallskip}
Channel & Wavelength Bands & FOV (Pixel size) \\
\noalign{\smallskip}
\hline
\noalign{\smallskip}
IRC-NIR & K,L,M+($1.25-2.5, 2-5.5\mu$m grism) & $10'$x$10'(1.4''/pixel)$ \\
IRC-MIR-S & $7,9,11\mu$m($5-10\mu$m grism) & $10'$x$10'(2.34''/pixel)$ \\
IRC-MIR-L & $15,20,25\mu$m($10-25\mu$m grism) & $10'$x$10'(2.34''/pixel)$ \\
\noalign{\smallskip}
\hline
\noalign{\smallskip}
\end{tabular}\\
\label{IRC}
\end{table*}

\subsection{The Infra-Red Camera (IRC)}

 The IRC is a wide-field imaging instrument ~\cite{mat98}. It consists of three independent camera systems. IRC-NIR (1.8-5 $\umu$m), MIR-S (5-12$\umu$m) and MIR-L (10-25$\umu$m) - see Table~\ref{IRC} ~\cite{wat00}. At the focus of these cameras are new large format infrared array detectors (NIR:512x412 InSb, MIR:256x256 Si:As IBC). For IRC-NIR, only a 412 x 412 portion of the InSb array is used to observe the 10\arcmin x 10\arcmin field, the remaining detector area (412x100) being used for a slit spectroscopy via a slit added to the entrance aperture. The advantages of such infrared observations from space are the negligible atmospheric absorption and low background emission (1$\sim$5 photons/s for NIR, and 100$\sim$1000 photons/s for MIR). These 3 channels will have a field of view of 10\arcmin x10\arcmin. The MIR-S channel shares the FOV with the NIR channel with light of wavelength shorter than 5$\umu$m being transmitted to the IRC-NIR and incident light of longer wavelength being reflected to the MIR-S channel, while the MIR-L-IRC FOV is separated by approximately 20\arcmin. At the aperture stop of each camera (i.e., the image position of the telescope primary mirror, $\approx$11$\sim$12mm in diameter), the observing wavelength is selected by a filter wheel. The filter bands are selected by rotating the filter wheels. The IRC-NIR filter wheel has 6 positions. 3 bandpass filters of Ge substrates (6mm thick), 2 dispersive elements and a blank mask. The 3 band pass filters of the MIR-S are made of ZnS substrates of 7mm thickness and the grism (used for spectroscopy) has 4.8 grooves per mm. The spare filter wheel position (TBD) will be another grism or a broad band filter. The MIR-L channel is located $25\arcmin$ away from the telescope axis and has 4 filters on CdTe substrates of 8mm thickness plus a grism (2.5 grooves/mm) and a spare as yet undecided position for possibly another grism or a broad band filter.

 In addition to the narrow-band photometric filter bands, there is also a blank position to measure the dark level, and positions for grisms. Thus the IRC also has the capability of low-dispersion spectroscopy by replacing the filters with the grisms attached to the filter wheels. This spectroscopic capability will be used to classify detected sources, to obtain source SED's and to roughly estimate their redshifts. There also exists one spare position on the filter wheel that is as yet undecided but could possibly be utilized as a another grism or a broad band filter. 

The IRC will be operated under pointing mode where the satellite attitude is fixed for $\approx$10 minutes in order to observe one target direction on the sky, the typical exposure time per pointing being 500 seconds in total. Such pointing observations can be made 1-3 times per revolution of the satellite. For the in-flight calibration of detector response and flat-fielding, a light source with uniform intensity is also included. Illumination of a rough light-scattering surface of a blank position on the filter wheel facing the detector array before and after each pointing observation will establish the flat-fielding with about $1\%$ accuracy. However, for the MIR cameras, a flat-field accuracy better than $0.1\%$ is required. Therefore each pointing observation ($\approx$10 minutes) is divided into 8$\sim$10 independent observations in units of the exposure time of the IRC-NIR ($\approx$ 62s, 60s for a dark sky and 2s exposure for observation of bright sources in the case of the NIR-IRC and 6x10 + 2x1s for the MIR-IRC \cite{wat00} ), with the telescope position on the sky being moved by 5\arcsec$\sim$30\arcsec from the origin of the observing field among the sub-observations.

\section{Mid-infrared Model Parameters}\label{sec:model}

Galaxy source counts are simulated by using an extended and improved evolution of the Pearson and Rowan-Robinson model ~\cite{cpp96}. The model utilizes a 4 component parameterization based on the IR colours of IRAS galaxies ~\cite{RR89}. The model consists of a normal galaxy (IR cirrus like) component defined by cool 100/60$\umu$m colours, a starburst component defined by warm 100/60$\umu$m colours extending to a high luminosity tail ($L_{60\mu m} > \sim 10^{12} L\sun $) in the IR luminosity function to model the ultraluminous infrared galaxy population ~\cite{sand96} and an AGN (Seyfert/QSO 3-30$\umu$m dust torii) component ~\cite{RR95}. Components are distinguished on a basis of SED and luminosity class. The composite cool and warm 60$\umu$m luminosity functions of Saunders et al. ~\shortcite{saun90} are used to represent the normal, starburst and ULIG galaxies respectively. The {\it cool} galaxies are defined in $\nu L_{\nu}$ from $lgL_{60\mu m} = 6-11.5L\sun $, the starbursts from $lgL_{60\mu m} = 6-12L\sun $ and the ULIG component from $lgL_{60\mu m} > 12L\sun $. The AGN luminosity function is defined from $lgL_{12\mu m} = 8-14L\sun $ at 12$\umu$m ~\cite{rush93} using the luminosity function of Lawrence et al. ~\shortcite{law86}. To shift the luminosity function from the wavelength at which the luminosity function is defined, $\lambda _{LF} $, to the observation wavelength $\lambda _{obs} $, the ratio $L(\lambda _{obs})/L(\lambda _{LF} )$ is obtained via model template spectra. 

K-corrections are calculated using model Spectral Energy Distribution (SED's) templates for each galaxy population. The original model SED's of Pearson \& Rowan-Robinson ~\shortcite{cpp96} consisted of normal, starburst, AGN and ultra/hyperluminous galaxy spectra and were found to provide a good fit to IRAS data. However much progress has been made both in the fields of observation with the Infrared Space Observatory (ISO) and radiative transfer models. Of particular significance for predictions in the mid-infrared region is the inclusion of the so called unidentified infra-red bands probably manifested as Polycyclic Aromatic Hydrocarbons (PAH's ~\cite{pug89}), into the model spectra. Such features, prominent in the 6-12$\umu$m range would be expected to affect the mid-IR source counts to some extent~\cite{xu98}. Furthermore, in the wake of the many observations by ISO (~\cite{lu97}, ~\cite{boul96},~\cite{vig96}), it has become starkly apparent that most galaxies are hosts to such features, with equivalent widths of as much as 10$\umu$m, comparable to that of the ISOCAM MIR filter bandpasses (3.5$\umu$m \& 6$\umu$m for the LW2(5-8.5$\umu$m) filter and LW3(12-18$\umu$m) filters respectively).

\begin{figure}
\centering
\centerline{
\psfig{figure=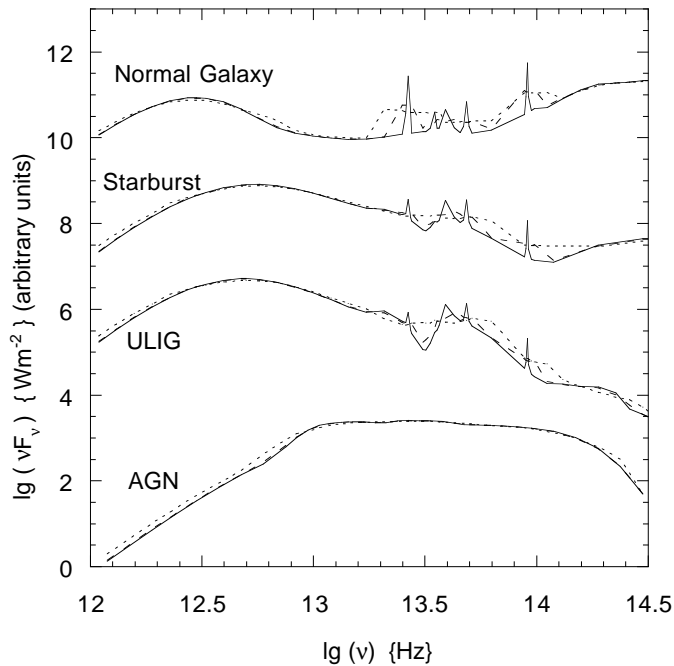,height=10cm}
}
\caption{ The model SED templates used for the source count models. Also shown is the effect of smoothing the original template SEDs ({\it solid line \,}) by the filter bandwidth at each wavelength. In the case of the narrow band filters $\Delta \lambda / \lambda = \Delta \nu / \nu = 0.2$ ({\it dashed line \,}). If the wide band filter option is utilized, $\Delta \lambda / \lambda = \Delta \nu / \nu = 0.67$ ({\it dotted line \,})
\label{smooth}}
\end{figure}

The template SED's for the starburst \& Ultraluminous {\it warm} components and the normal {\it cool} galaxy component are taken from the new models of Efstathiou, Rowan-Robinson \& Siebenmorgen ~\shortcite{esf001} and Efstathiou \& Siebenmorgen ~\shortcite{esf002} respectively, while the AGN SED template is that of Rowan-Robinson ~\shortcite{RR95}. The new starburst galaxy SED's are based on the evolution of optically thick giant molecular clouds, centrally illuminated by massive stars. The starburst models are defined by {\it t\ }, the age of the starburst in Myrs; $\tau$, the initial optical depth (in {\it V\ }) of the molecular clouds; $\chi$, the ratio of radiation field to the local solar neighborhood. The starburst SED can be considered M82 {\it like}, while the ULIG SED represents an ARP220 type galaxy (c.f. the more extreme hyperluminous galaxy IRAS F10214 ~\cite{RR932} that was used in the earlier models of Pearson \& Rowan-Robinson~\shortcite{cpp96} to represent the high end of the starburst luminosity function). Exponentially decaying $10^{7} \sim 10^{8}$yrs old starbursts reproduce well the IRAS colours of galaxies and provide good fits to the data for the popular starburst galaxy template M82 and to the recently ISO observed starburst galaxy NGC6090 ~\cite{acost96}. The new galaxy SED templates also incorporate the full range of PAH features ~\cite{sieb92} crucial for mid-infrared predications and not previously included in the earlier model SED's ~\cite{RR92}, ~\cite{RR931}. It should be noted that although the PAH features seem somewhat strong and sharply defined in these models, as long as the energy contained within the peaks is consistent with observations and the SEDs are smoothed and convolved with the filter response function, the general shape of the PAH features is not so critical (see fig.~\ref{smooth}).

We consider 2 evolutionary models. The first model may be considered a {\it safe, well established benchmark \ } for the survey predictions (we refer to this as the {\it Levol} (luminosity evolution) model. For this model, simple luminosity evolution of the form $L(z)=L(0)(1+z)^{3.1}$ to a redshift of $z=2$ is incorporated into the starburst, ULIG, and AGN components. This form of evolution has been shown to provide a good fit to the source counts over a wide range in wavelength ~\cite{RR99}, ~\cite{cpp96}, ~\cite{benn93},~\cite{boyl88} although additional evolution is almost certainly needed to explain the faintest 15$\umu$m ISOCAM \& 170$\umu$m ISOPHOT source counts ~\cite{elbaz99}, ~\cite{elbaz99b}, ~\cite{dole00}, \cite{biviano99}, \cite{metcalfe00}, \cite{altieri99} and the SCUBA sub-mm data ~\cite{hugh98},~\cite{smail97},~\cite{blain99},~\cite{barg99}.

The second model incorporates more extreme evolution. In this model, the starburst and AGN evolve {\it similarly} to the {\it Levol} model with a power law of $L(z)=L(0)(1+z)^{3.1}$, however, in addition, the high luminosity end of the starburst luminosity function (ULIG population) evolves not only in luminosity but also in number exponentially. We will refer to this model as the {\it ULIG} model. Although locally ($\approx z<0.1$), ULIGs have a space density approximately similar to optically selected QSOs, analysis of the IRAS Faint Source catalogue has hinted at much stronger (density) evolution at higher redshifts (to $z \approx 1$) of the form $(1+z)^{g}$, with $g \approx 7.6\pm 3.2$ ~\cite{kim98}. Similar strong evolution ($g \approx 10$) is also required to fit the deep 15$\umu$m ISO source counts ~\cite{dole00}. Using such strong evolution the source counts from sub-mm to NIR wavelengths can be fitted by one consistent model (see Pearson ~\shortcite{cpp002} for detailed description of these new models). 

\begin{figure}
\centering
\centerline{
\psfig{figure=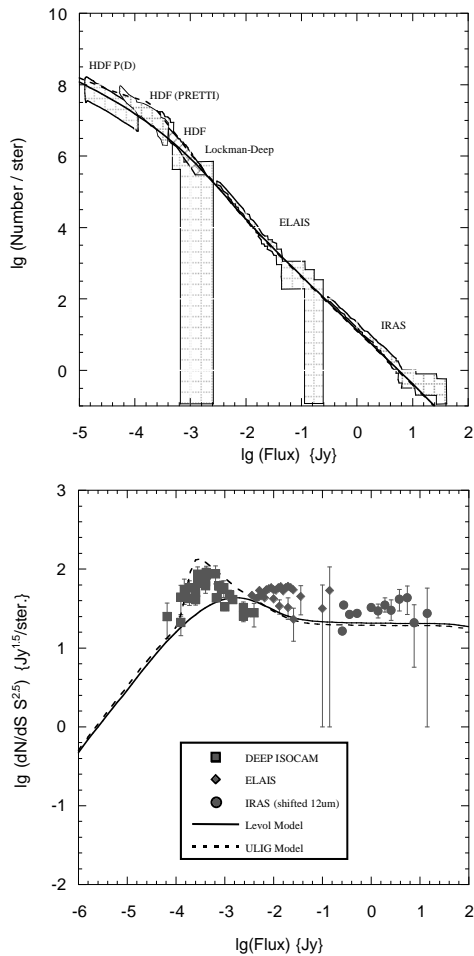,height=14cm}
}
\caption{Source counts at 15$\umu$m for the {\it Levol} \& {\it ULIG} models compared with the source counts as observed by ISO. {\it Top} - Integral source counts compared with observations from Hubble Deep Field (P(D) analysis - Oliver et al. (1997), HDF counts derived using the PRETI method - Aussel et al. (1999), Lockman Hole - Elbaz et al. (1998), ELAIS - Serjeant et al. (2000) and shifted IRAS counts. {\it Bottom} - Normalized differential counts compared with observations - Elbaz (1999b), Serjeant et al. (2000).
\label{15um}}
\end{figure}

The strongest constraint on the evolution in the mid-infrared are the source counts at 15$\umu$m ~\cite{elbaz99}, where the counts have been measured over 4 orders of magnitude in flux. These counts exhibit a super Euclidean slope at fluxes of around $1mJy$, flattening at fainter fluxes of $0.4mJy$ and are best shown by plotting the differential source counts. Fig.~\ref{15um} shows the {\it ULIG} and {\it Levol} model fits to the 15$\umu$m source counts. As can be seen from the figure, although the {\it Levol} model provides a very good general fit to source counts over a wide range in wavelength (and should therefore not be discarded out of hand by any means (see Serjeant et al. ~\shortcite{serjeant00} for a detailed comparison of the Pearson \& Rowan-Robinson model with ISO 15 \& 6.7$\umu$m observations.), it cannot however, reproduce the detail of the dips and peaks observed in the ISO counts. However for a general investigation and prediction of the capabilities of ASTRO-F, this simple evolution will suffice.

Note that although an E/S0 component is not strictly included in the models, the strong evolution in the starburst component adequately compensates for this absence ~\cite{cpp96}. Furthermore any significant contribution from such a population decreases rapidly towards longer wavelengths. The model assumes $\Omega=0.1, H_o = 50kms^{-1}Mpc^{-1}$.

\section{Mid-infrared surveys with the IRC}\label{sec:survey}

\subsection{The effect of smoothing the SEDs by the filter bandwidths}

With the IRC-MIR on ASTRO-F, mid-infrared surveys will be possible over a wide range of wavelengths from 6-25$\umu$m. With such a wide range of options and combinations available for mid-infrared surveying, it is of paramount importance that the filters be used in the most efficient way, in order to, for example, detect the most sources or to discriminate between galaxy types or redshift domains. Due to the dominance of the PAH features in the mid-infrared, the filter bandwidth will have a significant effect on the results. A previous, initial estimation of probable survey outcomes made by Pearson ~\shortcite{cpp001} assumed monochromatic survey filter wavelengths and hence suffered from prominent spikes in the N-z distribution of galaxies due to the effect of the PAH features in the mid-IR (see fig.~\ref{nzdeep} (a)). The narrow band filters have a resolution, $\Delta \lambda / \lambda = \Delta \nu / \nu = 0.2$. Therefore the template SEDs are convolved with the filter bandwidths at each wavelength to produce a ``smoothed'' SED template. Such smoothing provides a much more realistic picture of the mid-infrared SED of galaxies as will be seen by ASTRO-F. The smoothed normal and starburst galaxy SED's are shown in Fig.~\ref{smooth}. Note that the effect of smoothing the AGN SED is negligible due to the fact that the mid-infrared broadband features are not present in the AGN spectra having been suppressed / destroyed by the intense radiation field in the AGN environment~\cite{lutz97},~\cite{roache91},~\cite{genzel98}. The most obvious difference between the smoothed and unsmoothed SED's is the significant reduction of the amplitude of the prominent PAH features which should have the effect of {\it smoothing} the galaxy {\it N-z} distributions. 

In the following subsections, surveys with this narrow band filter configuration are considered.

\subsection{Survey strategies and detection limits}

\begin{figure*}
\centering
\centerline{
\psfig{figure=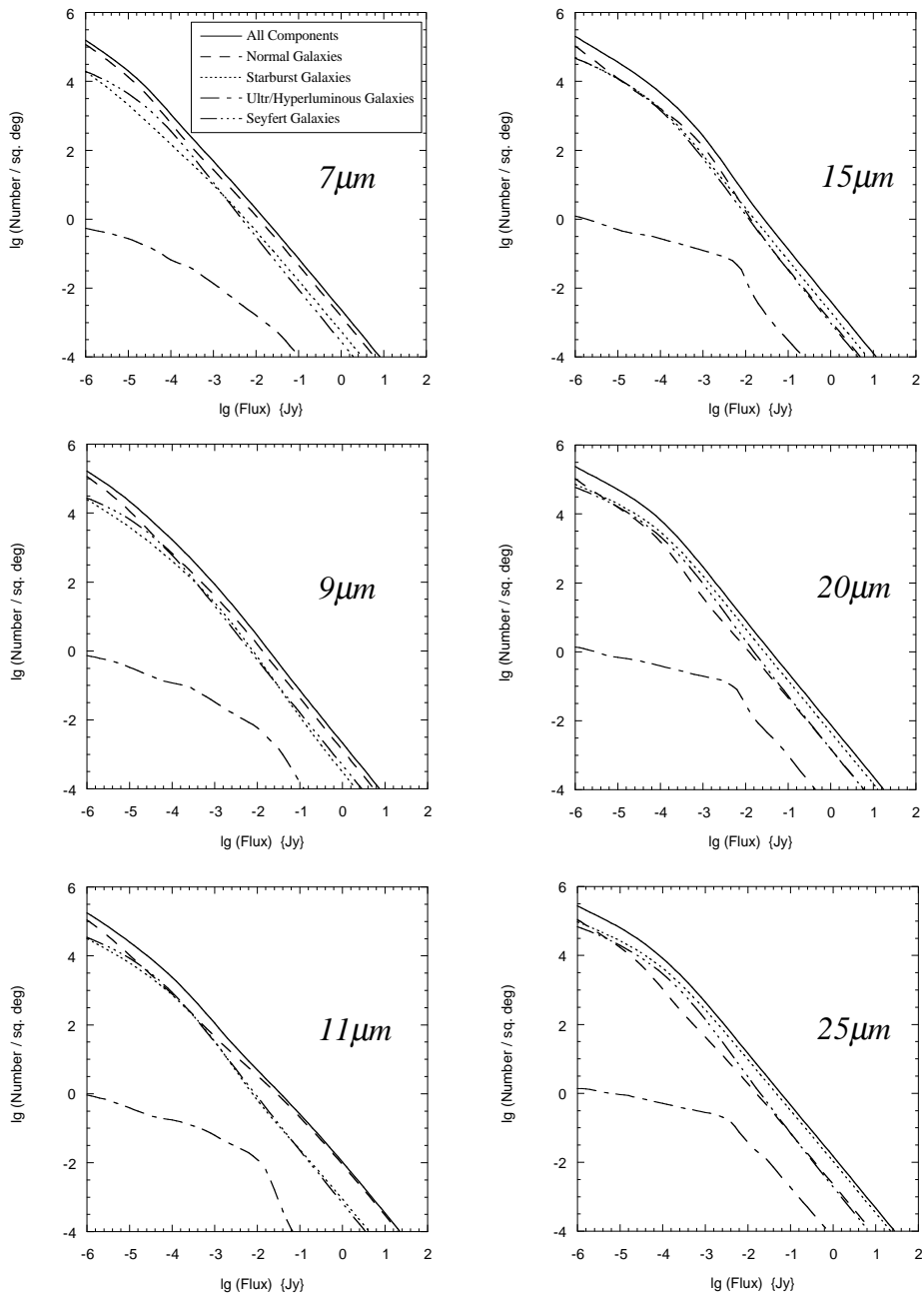,height=18cm}
}
\caption{Integral galaxy source counts, assuming the {\it Levol} model, at 7, 9, 11, 15, 20 \& 25$\umu$m respectively, corresponding to the IRC-MIR-S and IRC-MIR-L filter wavelengths. Total source counts are shown ({\it solid line\,}) with normal ({\it dash}), starburst ({\it dot\,}), ultraluminous galaxy ({\it dash dot\,}) and AGN components ({\it dot-dot-dot-dash\,}). There is a gradual trend of increasing starburst galaxy contribution towards the longer wavelengths.
\label{intctsL}}
\end{figure*}

\begin{figure*}
\centering
\centerline{
\psfig{figure=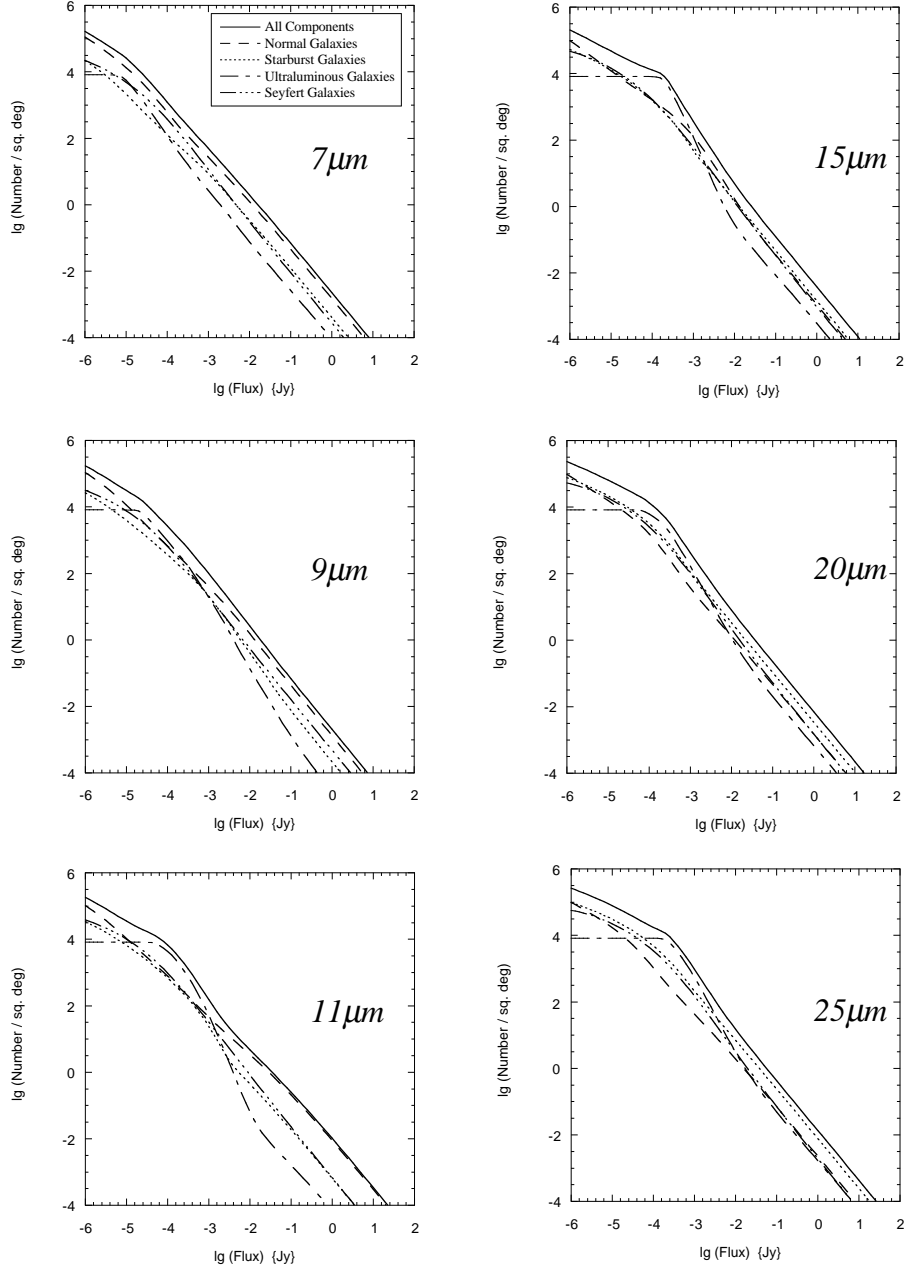,height=18cm}
}
\caption{Integral galaxy source counts, assuming the {\it ULIG} model, at 7, 9, 11, 15, 20 \& 25$\umu$m respectively, corresponding to the IRC-MIR-S and IRC-MIR-L filter wavelengths. Total source counts are shown ({\it solid line\,}) with normal ({\it dash\,}), starburst ({\it dot}), ultraluminous galaxy ({\it dash dot\,}) and AGN components ({\it dot-dot-dot-dash\,}). Note the predominant bump produced by the strongly evolving ULIG component as compared to the relative scarceness of these sources in the {\it Levol} model scenario.
\label{intctsU}}
\end{figure*}

Mid-infrared source counts from 7-25$\umu$m, corresponding to the IRC-MIR-S and IRC-MIR-L filter wavelengths are shown in detail in fig.~\ref{intctsL} and fig.~\ref{intctsU} for the {\it Levol} and {\it ULIG} models respectively. In general there is an increasing contribution from the starburst (and AGN) component towards longer wavelengths. In the case of the {\it ULIG} model the source counts at all wavelengths exhibit a hump at the flux where the ULIG component peaks and then consequently flattens.

\begin{table*}
\begin{minipage}{110mm}
\caption{ASTRO-F Narrow Band Filter Survey Detection Limits.}
\label{limits1}
\begin{tabular}{@{}lcccc}
wavelength & \multicolumn{4}{c}{Sensitivity limit ($\mu$Jy)} \\
   &  shallow survey &  deep survey &  \multicolumn{2}{c} {source confusion limit} \\
   & (single pointing) & ($\approx$ 20 pointings) & Levol model & ULIG model \\
\hline
$7\mu$m & $20$ & $4$ & $2.7$ & $3.3$ \\
$9\mu$m & $31$ & $7$ & $5.2$ & $7.8$ \\
$11\mu$m & $53$ & $12$ & $9.2$ & $14.5$ \\
$15\mu$m & $85$ & $19$ & $29.9$ & $76$ \\
$20\mu$m & $120$ & $27$ & $83.4$ & $158$ \\
$25\mu$m & $150$ & $34$ & $149.3$ & $363$ \\
\hline
\end{tabular}
\end{minipage}
\end{table*}
\begin{table*}
\begin{minipage}{110mm}
\caption{Comparison of ISO and ASTRO-F 15$\mu$m Surveys.}
\label{surveys}
\begin{tabular}{@{}llll}
Survey Name & Reference & Area & Sensitivity \\
\hline
ISO Lensed & \cite{biviano99} & $56\arcmin^2$ & $0.07mJy$ \\
HDF & \cite{serjeant97} & $5\arcmin^2$ & $0.2mJy$ \\
CAM shallow & \cite{elbaz99} & $0.41^{\sq}$ & $0.8mJy$ \\
CAM deep & \cite{elbaz99} & $0.3^{\sq}$ & $0.5mJy$ \\
ASTRO-F deep NEP & - & $3200\arcmin^2$ & $0.04mJy$ \\
ELAIS & \cite{RR99} & $12^{\sq}$ & $2mJy$ \\
ASTRO-F shallow & - & $18^{\sq}$ & $0.085mJy$ \\
\hline
\end{tabular}
\end{minipage}
\end{table*}

Table~\ref{limits1} shows the current expected sensitivities for a single pointing ($5\sigma$) using the narrow band filters on the IRC-MIR detectors. The confusion limit due to background sources is calculated from the source counts and is also tabulated in table~\ref{limits1}. For any space telescope, the confusion limit at mid-infrared wavelengths will be lower than in the far-infrared (and lie at higher redshift) due to the higher spatial resolution at the shorter wavelengths ($ \sim \lambda /D$).  The confusion limits are calculated from the source counts assuming the classical confusion criteria of a source density of 1 source per 40 beams of the observing instrument, where the beam diameter is given by $d=1.2 \lambda /D$, where $D$ is the telescope diameter (70cm). Here we calculate the source confusion assuming a beam diameter equivalent to the FWHM of the Airy disc, $d=1.2 \lambda /D$ such that the beam diameter at 15$\umu$m is 5.3\arcsec corresponding to a source confusion density of $\approx$14660 sources. The corresponding limiting flux will of course be model dependent.

We investigate 2 general strategies for possible mid-infrared surveys with ASTRO-F.

 The first is an ultra-deep survey covering $\approx 3200 sq. arcmin.$ around the North Ecliptic Pole (NEP), making 20 pointings per pixel giving an effective gain of $\sqrt{20}$ over the base sensitivity (given in table~\ref{limits1}). The NEP is chosen due to the fact that ASTRO-F will be in a sun-synchronous polar orbit with the sunshield permanently facing inward towards the Sun and the telescope direction always approximately on the plane perpendicular to the direction of the Sun (i.e., toward the pole). Thus a minimal amount of survey planning and telescope moving is required. 

The second survey strategy is a wide field shallow survey with a single pointing per pixel covering approximately $20$x$3200sq. arcmin. \approx 18^{\sq}$. The area covered by the shallow wide field survey would be larger, and the sensitivity $\sim 20$ times deeper at 15$\umu$m than that of the European Large Area ISO Survey (ELAIS~\cite{oliver00b} which impressively covered $\sim 12^{\sq}$ at 15$\umu$m to $\approx 2mJy$ (table~\ref{surveys}). 

\begin{figure}
\centering
\centerline{
\psfig{figure=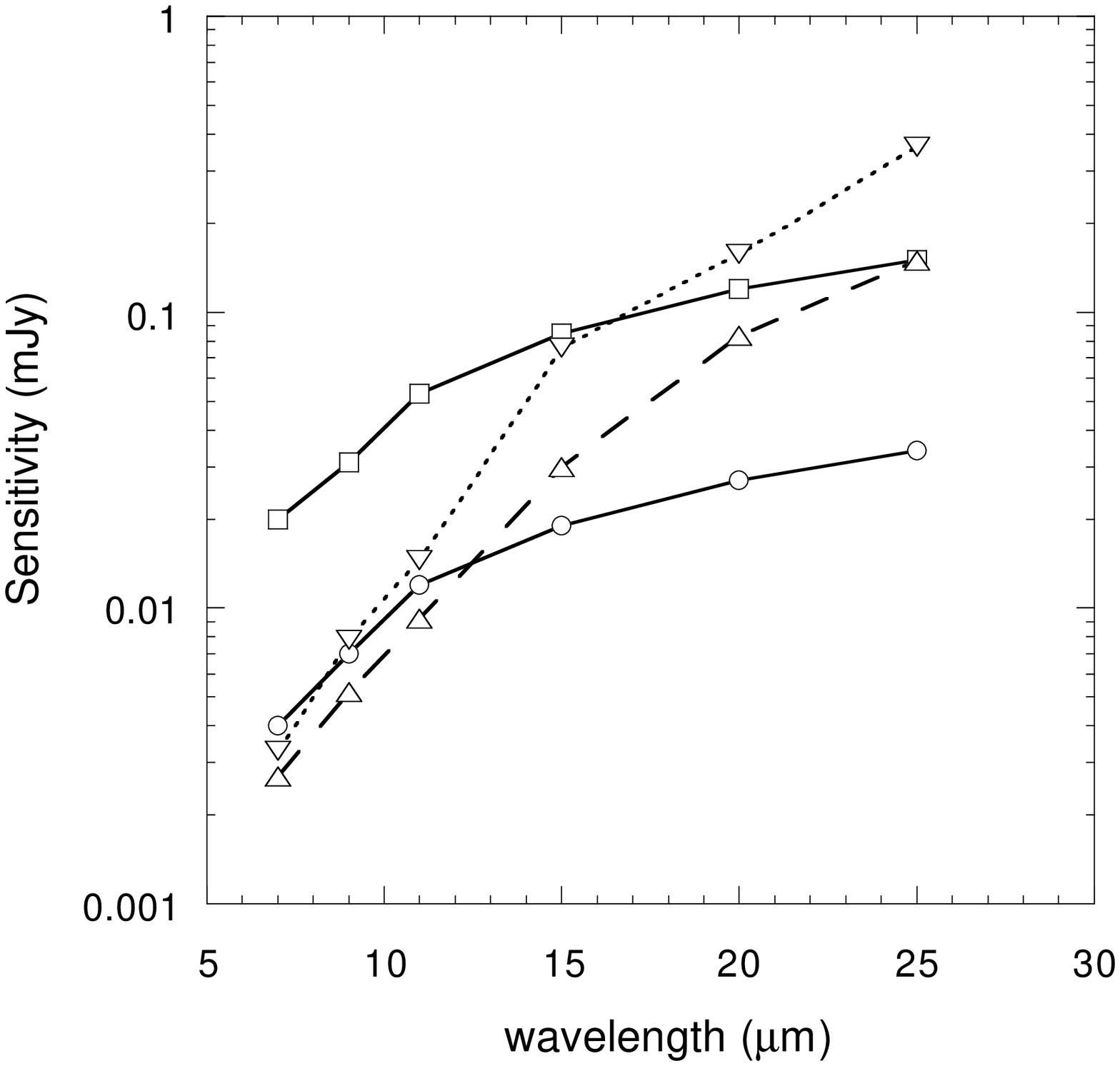,height=10cm}
}
\caption{Survey sensitivities at 7, 9, 11, 15, 20 \& 25$\umu$m shown for the proposed 3200sq.arcmin. deep NEP survey ({\it circles}), the 18sq.deg. shallow (single pointing) survey ({\it squares}) and the corresponding source confusion limits predicted by the {\it Levol} ({\it up triangles\,}) and {\it ULIG} models ({\it down triangles\,}). For the deep survey, in general the shorter wavelength observations with the IRC-MIR-S remain above ({\it Levol} model) or close to ({\it ULIG} model) the source confusion limit. However the longer wavelength observations with the IRC-MIR-L are expected to be strongly constrained by the confusion due to faint sources. For the wide, shallow survey the sensitivities attainable at longer wavelengths (MIR-L) depend critically on the evolutionary models.
\label{confuse}}
\end{figure}

From table~\ref{limits1}, it is apparent that such mid-infrared surveys may well be severely constrained by source confusion. In fig.~\ref{confuse} a comparison is made between the survey sensitivities expected for the 3200sq.arcmin. deep NEP survey, the 18sq.deg. shallow (single pointing) survey and the corresponding source confusion limits predicted by the {\it Levol} and {\it ULIG} models. Deep surveys with the MIR-S instrument are almost on the confusion limit due to background sources (the optimal position required!), however at the longer wavelengths corresponding to the MIR-L, the surveys is expected to be heavily contaminated by the background source confusion. In the case of a wide field shallow survey, there is a greater dependence on the evolutionary model assumed. Assuming the {\it Levol} model, all bands are above or on the background confusion limit. However, if the {\it ULIG} model is assumed, then the MIR-L bands still suffer contamination to some extent in the shallow survey as well as the deep NEP survey. Thus, it is apparent that any future surveys in the mid-infrared will be severely limit by source confusion. Even in the case of ISO, observations in the Hubble deep Field at 15$\umu$m down to 255$\umu$Jy were already reaching the confusion limit ~\cite{oliver97}. For ASTRO-F, the problem of source confusion and subsequent source extraction will be somewhat alleviated by the fact that for the deep NEP survey the observations with the IRC-MIR-S, IRC-MIR-L \& IRC-NIR instruments overlap allowing multiple observations at different wavelengths.

The main limitations of the ISO deep surveys were effects caused by flat-fielding noise, glitches and transients \cite{aussel99}. Although detailed modelling of these effects and their possible influence on the ASTRO-F IRC surveys is being undertaken, it should be noted that the IR arrays used in the IRC-MIR channels utilize the IBC (impurity band conduction) structure that is superior to the bulk photoconductive detection arrays used in ISOCAM, especially with respect to transient behavior which will enable us to obtain better flat fielding. Furthermore, please note that at each pointing we take 6x(7$\sim$10)=42-60 frames with micro-scanning towards the same field of sky hence for 20 pointings the real redundancy is not 20 but 20x(42-60)=840-1200 \cite{wat00}.

\subsection{The 3200sq.arcmin. deep NEP MIR survey}

The 3200sq.arcmin. survey would take an extremely deep (32 different IRC pointings each 20 pointings deep) observation around a $\sim$67\arcmin diameter doughnut of width $\sim$10\arcmin (i.e. each pointing has an area of 10$\times$10sq. arcmin, the field of view of the IRC) around the NEP. The aim of this survey would be to image as deep as possible in all mid-infrared bands of the IRC. 

\begin{figure*}
\centering
\centerline{
\psfig{figure=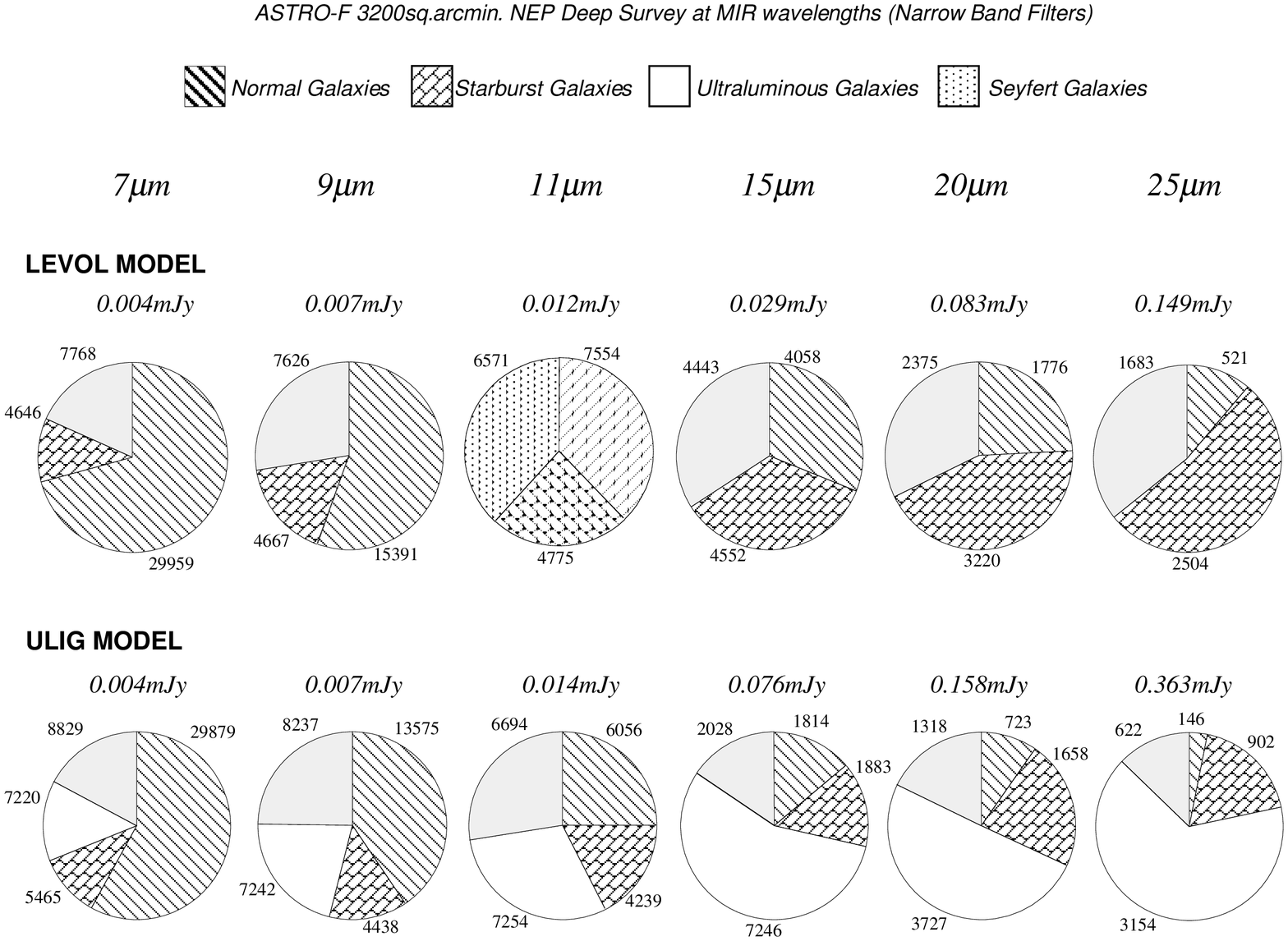,height=11cm}
}
\caption{Predicted numbers of galaxies for ASTRO-F deep 3200sq.arcmin. mid-IR survey. Predicted numbers and relative proportions of galaxies for both the Levol model ({\it top}) and ULIG model ({\it bottom}) are shown for the respective survey sensitivity limits (in $mJy$) given in table~\ref{limits1}. Note that if the survey sensitivity is source confusion limited then the sensitivity limit is dependent on the evolutionary model. The ultraluminous galaxy component for the {\it Levol} model is not included in the figures as their numbers are negligible.
\label{pie1}}
\end{figure*}

In  fig.~\ref{pie1} the predicted results for the proposed 3200sq.arcmin. Deep NEP MIR survey are shown as a function of object class (normal, starburst, ULIG and AGN) for both the {\it Levol} and {\it ULIG} evolutionary models. Note that the ultraluminous galaxy component for {\it Levol} model is not included as the expected number of sources are negligible (of the order of $<$ unity for the survey area). In the {\it Levol} evolutionary model, such objects may be serendipitously discovered and would be expected to be very rare objects, such as for example, the hyperluminous galaxy IRAS F10214+4724 ~\cite{RR91}, found at z=2.286 on the limit of the IRAS Faint Source Catalogue. The ASTRO-F deep mid-IR survey is not expected to find such objects since shallow, wide area strategies at longer wavelengths are more efficient for this purpose ~\cite{cpp96}. 

Of the order of 20,000-30,000 sources would be detected with the IRC-MIR-S instrument (7-11$\umu$m), the majority of which would be normal galaxies that have relatively strong K-corrections as the near-infrared/optical emission is sampled out to higher redshifts (the 3.3$\umu$m UIB feature being sampled at z$\sim$1.1 and the K-band at a redshift of $\sim$ 2.2 for the MIR-S 7$\umu$m band). Assuming the {\it Levol} model, $\approx 4000$ starburst galaxies and 4000-7000AGN would be expected in all bands except the longest(20-25$\umu$m) where the confusion limit becomes severe. In general the trend is towards an increasing fraction of starburst galaxies to normal galaxies towards longer wavelengths. The AGN component maintains an almost constant fraction of the total observed sources over the entire 7-25$\umu$m wavelength range. 

\begin{figure*}
\centering
\centerline{
\psfig{figure=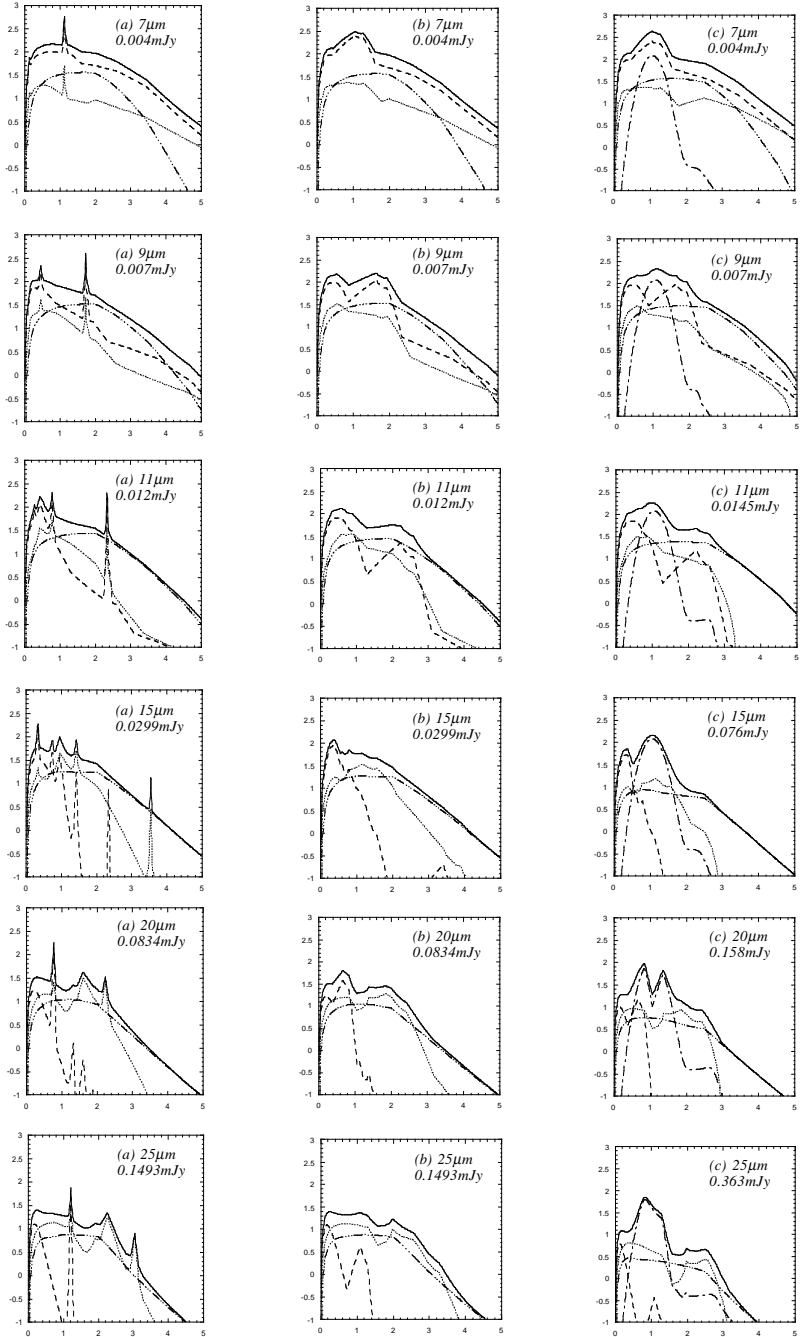,height=18cm}
}
\caption{N-z distribution of sources predicted for a deep survey with ASTRO-F around the NEP covering 3200sq.arcmin with survey sensitivities from tab.~\ref{limits1}. Vertical axes are ({\it lg(Number of sources / survey area (sq.deg.))\,}, horizontal axes are {\it redshift}. Panel a) N-z distributions using the unsmothed SEDs as spectral templates. Note the prominent spikes caused by the PAH features. Panel b) Results for {\it Levol} model. Panel c) Results for {\it ULIG} model. ({\it solid line\,}) - All Components, ({\it dashed line\,}) - Normal Galaxies, ({\it dotted line\,}) - Starburst Galaxies,({\it dot-dot-dot-dash line\,}) - AGN and ({\it dot-dash line\,}) - ULIG Galaxies. Note the ULIG contribution in the {\it Levol} model is too low to be included in the plots.
\label{nzdeep}}
\end{figure*}

Fig.~\ref{nzdeep} shows the number-redshift distributions for the deep NEP survey wavelengths as a function of galaxy component. Predictions for the {\it Levol (b)} and {\it ULIG (c)} models are shown. Also for comparison, the N-z distribution for the {\it Levol (a)} model using source spectra that have {\it not} been smoothed by the detector bandwidth are also shown. The prominent spikes produced as the PAH features pass through the bands, all but disappear when the SEDs are smoothed. This may have important consequences if the PAH features are hoped to be used as redshift indicators. 

The (smoothed) emission from the 3.3$\umu$m feature enhances the detectability of the normal galaxy component (and to a similar extent the starburst component) out to redshifts of $\sim$1-2.5 from 7-11$\umu$m respectively. In fact as much as 60$\%$ of the normal galaxies may lie at z$>$1 in the shortest 2 bands. In all bands, approximately 50$\%$ of the starburst galaxies lie at z$>$1 within which $\approx$10-20$\%$ may lie at z$>$2. Interestingly, the role of the mid-IR PAH features seems rather important enhancing the starburst galaxy population at redshifts of approximately 1 and 2 in the 15 and 20-25$\umu$m bands respectively as the mid-infrared region containing the 6-13$\umu$m {\it forest} of PAH features is redshifted into the respective observation windows.

Naturally, the {\it ULIG} model predicts starkly different results due to the inclusion of the strongly evolving new population. The fraction of ULIGs rises from around 10$\%$ in the shortest waveband to 65$\%$ in the longest. The number of ULIGs predicted remains approximately constant $\sim$7000 from 7-15$\umu$m, dropping to $\sim$3000 at the longest wavelengths due in part to the lower sensitivity of these bands. In the 7-15$\umu$m bands we are essentially seeing the entire ULIG population. The behavior of the other components being similar to the {\it Levol} model. Once again, significant contribution comes from the PAH features in the ULIG SED at the longer wavelengths. In all wavebands the contribution from the ULIG component peaks at about redshift 1 due to the form of the strong density evolution assumed by the model. Also prominent at 20$\umu$m is a double hump in the ULIG number redshift distribution. This feature is prominent due to the evolution incorporated into the ULIG model. In this scenario there is a peak and dip at a redshift $\approx$1 in the ULIG population corresponding to the redshift of the 9.7$\umu$m silicate absorption band observed at 20$\umu$m.

\subsection{The 18sq. deg. MIR shallow survey}

For the case of the ($\approx 18^{\sq}$) wide, shallow survey, the aim would be to cover as large an area as possible with the hope of serendipitously discovering new or rare mid-infrared objects. Of course, surveying such a large area in {\it all} IRC bands would be very time consuming with the same area having to be surveyed over 3 passes of the telescope, one for each of the filter wheel positions of the MIR-L and MIR-S. In reality, it may be that fewer filters (or a wide band filter) will be used. However, we present here the predictions for {\it all} IRC bands (7-25$\umu$m).

\begin{figure*}
\centering
\centerline{
\psfig{figure=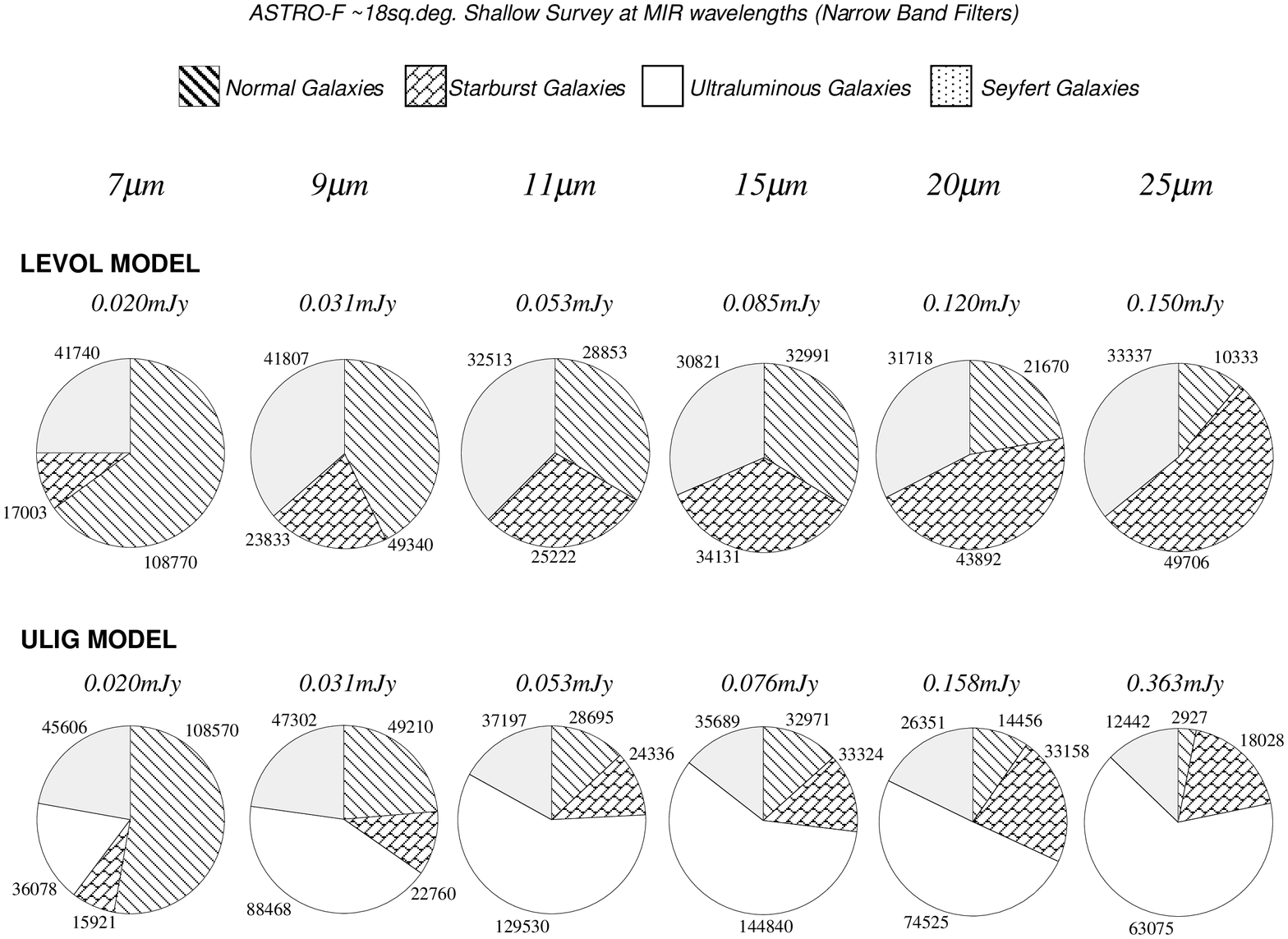,height=11cm}
}
\caption{Predicted number of galaxies for ASTRO-F shallow $\sim$18sq.deg. mid-IR survey. Numbers and relative proportions of galaxies for both the Levol model ({\it top}) and ULIG model ({\it bottom}) are shown for the respective survey sensitivity limits ({\it in mJy\,} - see table ~\ref{limits1}). Note that if the survey sensitivity is source confusion limited then the sensitivity limit is dependent on the evolutionary model. The ultraluminous galaxy component for the Levol model is not included in the figures as their numbers are negligible.
\label{pie2}}
\end{figure*}

In fig.~\ref{pie2} the predicted results for the wide, shallow survey are shown in the same form as the deep NEP survey. Note that again the ultraluminous galaxy component for {\it Levol} model is not included, although one of the aims of this survey would be the serendipitous discovery of such ultra/hyperluminous objects, with between a few to $\sim$10 of such objects predicted from 7-25$\umu$m respectively.

\begin{figure*}
\centering
\centerline{
\psfig{figure=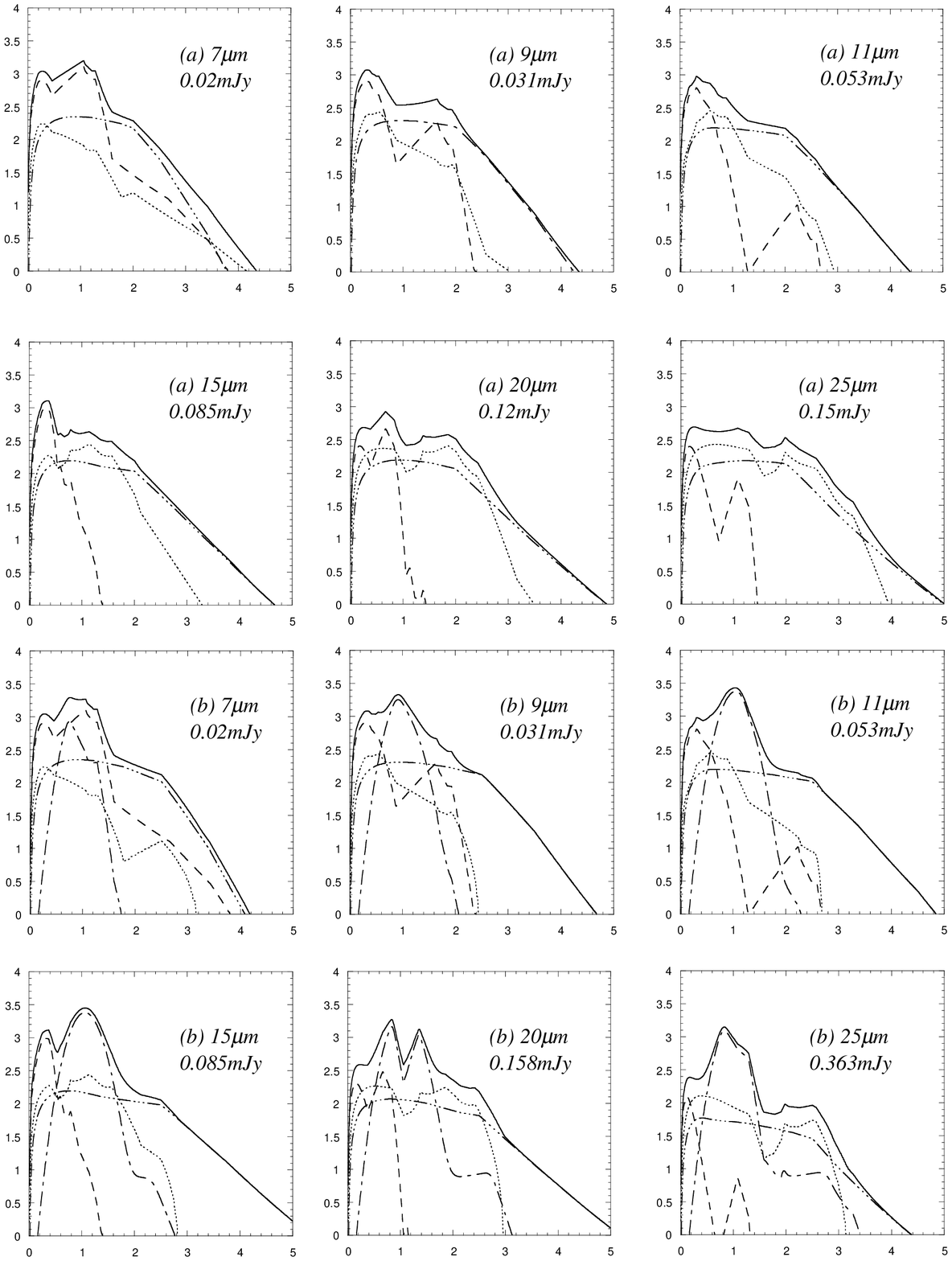,height=18cm}
}
\caption{N-z distribution of sources predicted for an 18sq.deg. mIR shallow survey with ASTRO-F with survey limits from table ~\ref{limits1}. Vertical axes are ({\it lg(Number of sources / survey area (sq.deg.))\,}, horizontal axes are {\it redshift}. Panel a) Results for {\it Levol} model. Panel b) Results for {\it ULIG} model.({\it solid line\,}) - All Components, ({\it dashed line\,}) - Normal Galaxies, ({\it dotted line\,}) - Starburst Galaxies,({\it dot-dot-dot-dash line\,}) - AGN and ({\it dot-dash line\,}) - ULIG Galaxies. Note the ULIG contribution in the {\it Levol} model is too low to be included in the plots.
\label{nzshallow}}
\end{figure*}

The number of normal, starburst and AGN galaxies predicted are similar for both evolutionary models with approximately 25,000 starburst galaxies and 35,000 AGN being detected in all bands and the proportion of normal galaxies gradually decreasing towards longer wavelengths. However, the major difference between the {\it Levol} and {\it ULIG} evolutionary models is the increasing contribution from and gradual dominance of the ULIG component in the {\it ULIG} model. This dominance is due to the strong evolution incorporated into the ULIG component peaking at a redshift of $\sim$1. Note that the results for the 20-25$\umu$m bands mirror those of the deep survey and are simply scaled by the increased survey area since both the deep and shallow surveys are source confusion limited. Fig.~\ref{nzshallow} presents the predictions for the number-redshift (N-z) distributions for the shallow survey. Naturally the N-z distributions for the deep and shallow surveys are quite similar in shape, the number of sources in the shallow survey being scaled by the larger area covered. In general the peaks produced by the PAH features at higher redshifts are weakened with respect to those at lower redshift due to the lower sensitivity of the shallow survey compared to the deep survey. Although in fact at the longest wavelengths (20-25$\umu$m) both the deep and shallow surveys would be source confusion limited. It may seem that at these wavelengths a deep survey may seem pointless, although it should be noted that P(D) analysis below the source confusion limit can also yield useful insight into the nature of the sources observed (e.g. ~\cite{mat00}).

\section{IRC Surveys with wide band filters}\label{sec:wide}

\subsection{The Wide Band Filter Option?}

There is still one spare position on the ASTRO-F-IRC filter wheels. One interesting option for this spare position would be a wide-band filter encompassing the entire wavelength range of the channel. i.e. a 6-12$\umu$m filter for the IRC-MIR-S and a 12-25$\umu$m filter for the IRC-MIR-L. The resolution for a central wavelength of 9$\umu$m would then be $\Delta \lambda / \lambda_{c} = 6/9 = 2/3$, thus giving a gain in sensitivity over the narrow filter configuration of $\sqrt{(2/3)/(1/5)} = \sqrt{10/3} \approx \sqrt{3}$. Although the gain in sensitivity is impressive, the associated loss of resolution and colour information must also be considered. Fig.~\ref{smooth} shows the effect of smoothing the model galaxy SEDs by the wide band filters.

The integral counts at the central wavelengths of 9$\umu$m and 18$\umu$m for the wide band filter configuration are not immediately visually different from the integral counts calculated using the narrow band filters and hence are not shown here. The sensitivities assumed for both the 3200sq.arcmin. deep NEP MIR survey and $\sim$18sq. deg. shallow MIR survey using the wide band filters are tabulated in table~\ref{limits2}. Also shown are the confusion limits due to faint background sources assuming the {\it Levol} and {\it ULIG} models respectively. From our previous investigation of the source confusion limits assuming the narrow band filter configuration (see table~\ref{limits1}), it is clearly seen that the severity of the source confusion increases with observing wavelength, the longer wavelengths becoming confused faster than the shorter wavelengths. Therefore, in order to calculate the source confusion limits for the wide band filters we assumed the longest wavelength in the band dominates the source confusion limits, (i.e.12$\umu$m for the 6-12$\umu$m channel and 25$\umu$m for the 12-25$\umu$m channel). This would be our worst case scenario for the survey sensitivity. From table~\ref{limits2}, it can be seen that there is little or no gain in the sensitivity of the deep survey over the shallow survey, both being basically constrained by the source confusion contamination.

Therefore, for the case of the deep 3200sq.arcmin. NEP MIR survey it would be prudent to use the narrow band filters as they remain (at least for IRC-MIR-S) above the confusion limit and there is little overhead or wasted time in changing the filters as the NEP could be observed easily many times. However for the larger area, shallow 18sq. deg. MIR survey, re-scanning the same area of sky 3 times (1 for each narrow band filter position on the IRC-MIR-S and MIR-L), may produce unreasonable overheads (this is without taking account of the additional scans for redundancy in the observations). Therefore the wide band filters would provide a faster, more economical alternative, with albeit, obvious disadvantages. For consistency, in this paper we discuss both deep and shallow survey wide band filter strategies.

\begin{table*}
\begin{minipage}{110mm}
\caption{ASTRO-F Wide Band Filter Configuration Survey Detection Limits.}
\label{limits2}
\begin{tabular}{@{}lcccc}
wavelength & \multicolumn{4}{c}{Sensitivity limit (mJy)} \\
   & shallow survey & deep survey &  \multicolumn{2}{c} {source confusion limit} \\
   & (single pointing) & ($\approx$ 20 pointings) & Levol model & ULIG model \\
\hline
$9\mu$m & $0.017$ & $0.004$ & $0.0108$ & $0.0172$ \\
$18\mu$m & $0.055$ & $0.012$ & $0.1210$ & $0.2767$ \\
\hline
\end{tabular}
\end{minipage}
\end{table*}

\subsection{ Wide Band Filter Survey Predictions}

In  table~\ref{widects} the predicted results for the deep 3200sq. arcmin. and shallow $\sim$18sq. deg. survey are shown.

Considering the deep 3200sq. arcmin. NEP survey, the {\it Levol} model predicts comparable numbers $\sim$20,000 sources, for both the wide 9$\umu$m band and the 9$\umu$m \& 11$\umu$m narrow bands. For the wide 9$\umu$m band, assuming the {\it ULIG} model, approximately 20,000 sources are predicted, compared to $\sim$25,000 and $\sim$30000 in the 9$\umu$m \& 11$\umu$m narrow bands respectively. This discrepancy between the models is due to the more severe constraint placed on the {\it ULIG} model by the confusion due to faint sources. At 18$\umu$m, where the constraint due to the source confusion is even more severe, both the {\it Levol} and {\it ULIG} models fall short of matching either the narrow band filter 15$\umu$m or 20$\umu$m predictions. Of course it should be kept in mind that the source confusion limits are set using the long wavelength edge of the wide bands.

However, considering the $\sim$18sq. deg. shallow survey, there is a shift in favour of the wider band filter configuration. Both the {\it Levol} and {\it ULIG} models predict higher numbers of sources in the 9$\umu$m wide band than in both the 9$\umu$m \& 11$\umu$m narrow bands. For example, assuming the {\it Levol} model, approximately 250,000 sources would be expected at 9$\umu$m with the wide band filter configuration, compared with the narrow band predictions of $\approx$100,000 at both 9$\umu$m \& 11$\umu$m. The {\it ULIG} model similarly predicts twice as many galaxies using the wide band filter configuration. In the case of the large area, shallow survey we are not integrating so deeply and therefore we do not become confusion limited so quickly. Therefore, the increased sensitivity gained in using the wide band filter configuration is not wasted. 

In the longer 18$\umu$m wide band, both the {\it Levol} and {\it ULIG} models predict of the order of 100,000 sources. As was the case before for the longer wavelength filters, confusion is again the limiting factor with the 18$\umu$m wide band configuration giving similar results as the longer wavelength filters (20-25$\umu$m) in the narrow band filter configuration.

\begin{figure}
\centering
\centerline{
\psfig{figure=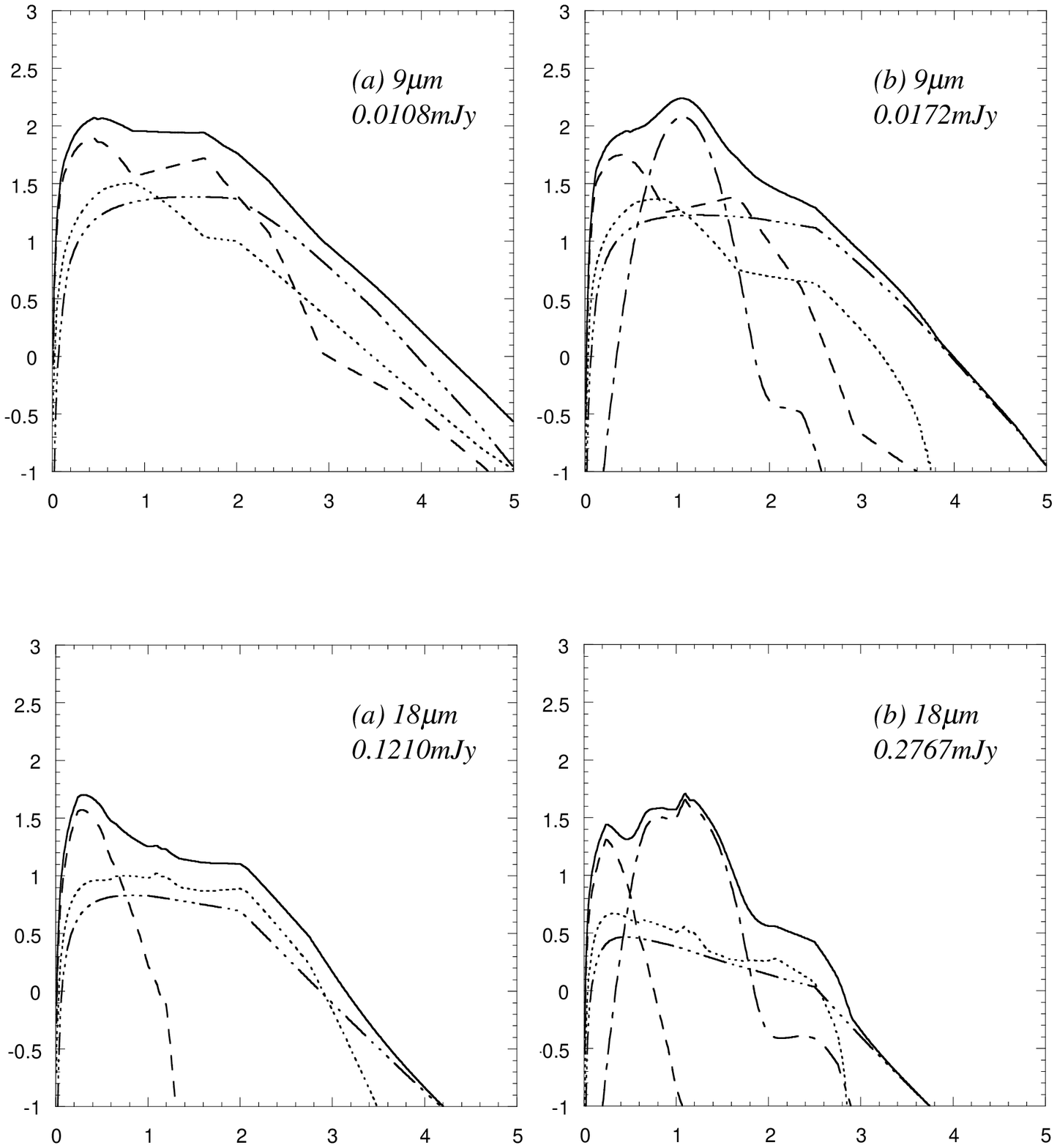,height=10cm}
}
\caption{Assuming the wide band filter configuration. 12$\umu$m and 18$\umu$m N-z distribution of sources predicted for a deep survey with ASTRO-F around the NEP covering 3200sq.arcmin with survey sensitivities from table~\ref{limits2}.. Vertical axes are ({\it lg(Number of sources / survey area (sq.deg.))\,}, horizontal axes are {\it redshift}. Panel a) Results for {\it Levol} model. Panel b) Results for {\it ULIG} model.({\it solid line\,}) - All Components, ({\it dashed line\,}) - Normal Galaxies, ({\it dotted line\,}) - Starburst Galaxies,({\it dot-dot-dot-dash line\,}) - AGN and ({\it dot-dash line\,}) - ULIG Galaxies. Note the ULIG contribution in the {\it Levol} model is too low to be included in the plots.
\label{nzwide}}
\end{figure}

In fig.~\ref{nzwide} the predicted number redshift distributions for the 3200sq.arcmin. deep NEP survey are shown for the {\it Levol} and {\it ULIG} models respectively. The corresponding results for the shallow survey are not plotted as they are essentially identical, being effectively scaled by the increased area of the survey. The main visible difference between the wide band N-z distributions and those of the narrow band is the further {\it smoothing out} of the influence of the infra-red bands (PAH features) and absorption features. Note that the 9.7$\umu$m absorption feature has all but disappeared from the longer wavelength N-z distributions.

\begin{table*}
\begin{minipage}{110mm}
\caption{Survey Predictions Assuming Wide Band Filter Configuration.}
\label{widects}
\begin{tabular}{@{}lcccccccc}
 & \multicolumn{4}{c}{Levol Model} & \multicolumn{4}{c}{ULIG Model} \\
Component   & \multicolumn{2}{c} {Deep} &  \multicolumn{2}{c} {Shallow} & \multicolumn{2}{c} {Deep} &  \multicolumn{2}{c} {Shallow}  \\
 & $9\mu$m & $18\mu$m & $9\mu$m & $18\mu$m & $9\mu$m & $18\mu$m & $9\mu$m & $18\mu$m \\
\hline
Normal Galaxies & $10460$ & $1710$ & $118350$ & $34220$ & $5800$ & $670$ & $115940$ & $13790$ \\
Starburst Galaxies & $4450$ & $2000$ & $59750$ & $40020$ & $3050$ & $700$ & $60960$ & $14030$ \\
Ultraluminous Galaxies & $0.3$ & $0.3$ & $5$ & $6$ & $7240$ & $3180$ & $144750$ & $63530$ \\
Seyfert Galaxies & $5470$ & $1380$ & $74450$ & $27640$ & $4240$ & $540$ & $84850$ & $10850$ \\
\hline
\end{tabular}
\end{minipage}
\end{table*}

\subsection{Galaxy Colours and Photometric Redshifts}

The mid and far infrared galaxies detected can be used to gain much information on both the galaxy type and evolution. It was on the basis of galaxy colours that the IRAS sources were divided into relatively distinct populations (cirrus, starburst, Seyfert)~\cite{RR89},~\cite{soif91}. In the mid-infrared, Fang et al. ~\shortcite{fang98} used the 25$\umu$m sample of Shupe, Fang \& Hacking ~\shortcite{shupe98} and the ISO observations of Lu et al.~\shortcite{lu97}, to calculate the 12/25$\umu$m colour-colour diagrams. In general, it was found that using a combination of 12/25$\umu$m colour and mid-infrared luminosity at 25$\umu$m, the different galaxy populations could be distinguished. For example, the luminous starburst galaxies were found to have systematically redder 12/25$\umu$m colours than those of AGN - a powerful discrimination tool. Results from the ISO-ELAIS survey have shown that stars and galaxies in such mid-IR surveys can also be easily discriminated by means of their mid-infrared colours ~\cite{crockett00}.

\begin{figure*}
\centering
\centerline{
\psfig{figure=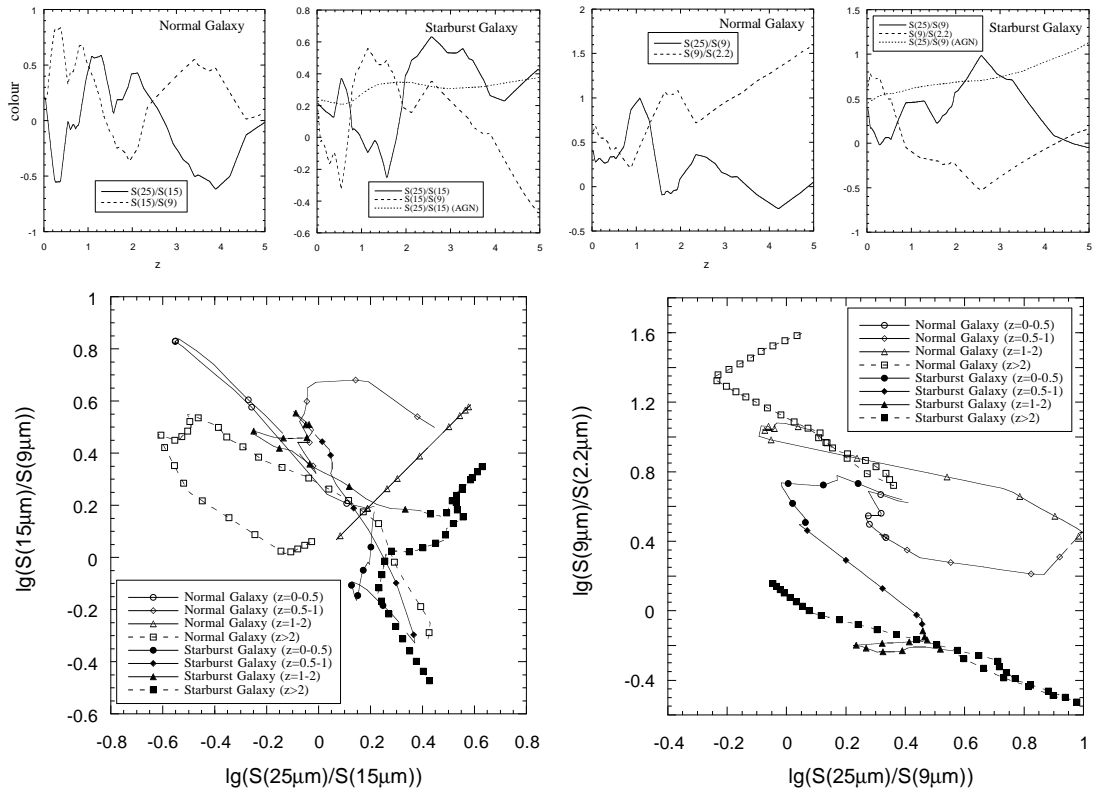,height=11cm}
}
\caption{Top panels - Colour-redshift distributions for normal and starburst galaxies (and AGN) assuming the narrow band filter configuration at 25, 15, 9 \& 2.2$\umu$m. Bottom panels - The 15/9$\umu$m - 25/15$\umu$m (left) and 9/2.2$\umu$m-25/9$\umu$m ({\it right}) colour-colour diagrams for normal ({\it unfilled markers\,}) and starburst ({\it filled markers\,}) galaxies. The results are divided into redshift regimes as described in the legend with each marker denoting 0.1 in redshift.
\label{colnarr}}
\end{figure*}

In fig. ~\ref{colnarr} we plot the 15/9$\umu$m and 25/15$\umu$m colour redshift relations for the normal (cirrus type) and starburst galaxy components assuming the {\it narrow band filter configuration}. Even though the model galaxy SEDs have been smoothed by the filter bandwidth, the contribution from the infrared unidentified bands (PAH features) severely contaminate the colours resulting in erratic switching of the dominant wavelength. This is further exemplified in the 15/9$\umu$m - 25/15$\umu$m colour-colour diagram. Although in general, the starburst galaxies tend to have warmer 25/15$\umu$m colours, there is much confusion and overlap between the normal and starburst populations. Furthermore, another motivation for measuring the colours of the observed sources would be to obtain some rough estimate of photometric redshifts. From the 15/9$\umu$m - 25/15$\umu$m colour-colour diagram, it can easily be seen that such estimations may be extremely difficult.

However, by combining measurements with the IRC-MIR-L (25$\umu$m), IRC-MIR-S (9$\umu$m) and IRC-NIR (2.2$\umu$m) channels, a better distinction can be achieved between both the galaxy populations and their photometric redshifts. Fig ~\ref{colnarr} also shows the 25/9$\umu$m and 9/2.2$\umu$m colour redshift relations and the corresponding 9/2.2$\umu$m-25/9$\umu$m colour-colour diagram. Here we find a better distinction as the introduction of the K-band channel effectively introduces a bimodal segregation into the starburst and normal galaxy populations (especially at the highest redshifts) with the normal galaxies generally having cooler 9/2.2$\umu$m colours than all but the lowest redshift starburst galaxies. However there is still substantial {\it contamination} as the unidentified band {\it PAH forest} is redshifted into the observing bands.

\begin{figure}
\centering
\centerline{
\psfig{figure=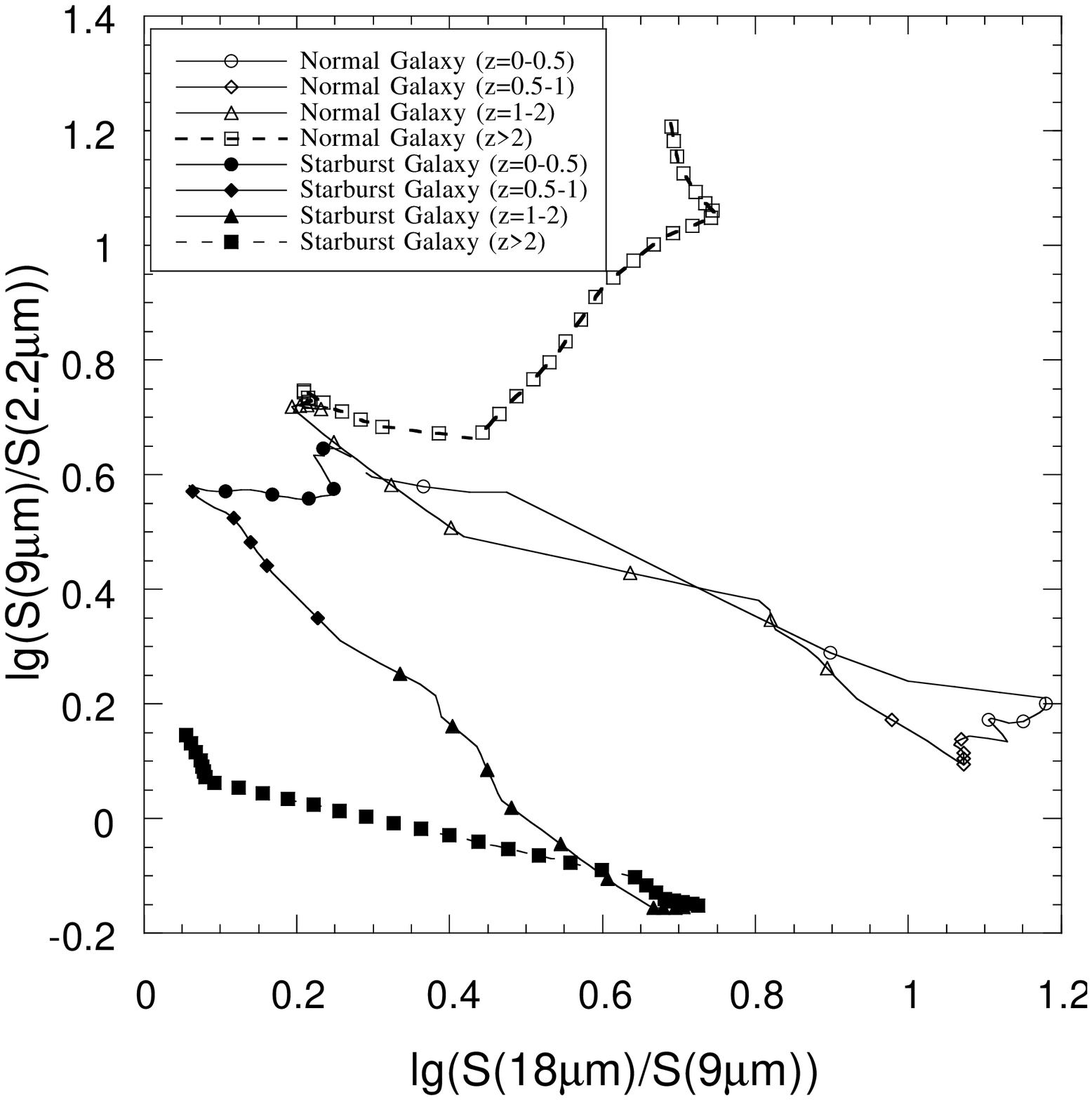,height=10cm}
}
\caption{The 9/2.2$\umu$m - 18/9$\umu$m colour-colour diagrams for normal ({\it unfilled markers\,}) and starburst ({\it filled markers\,}) galaxies assuming the wide band filter configuration. The results are divided into redshift regimes as described in the legend with each marker denoting 0.1 in redshift. Note that although in practice the variance in the individual galaxy colours will cause a {\it smudging } in the colour-colour plane, the substantial separation between the 2 populations should still remain. 
\label{colwide}}
\end{figure}

An interesting alternative is to use the wide band filter configuration (combined with one of the IRC-NIR channels e.g. K-band 2.2$\umu$m) to obtain colour-colour distributions in the hope that the contamination due to the unidentified bands will be further {\it washed out}. In fig.~\ref{colwide} the 9/2.2$\umu$m-18/9$\umu$m colour-colour diagram is plotted. Here we see a striking bimodal distinction between the normal and starburst galaxy components. In general the lower redshift normal galaxies have cooler colours than their starburst counterparts with the higher redshift areas again being completely distinct. It should also be noted from fig.~\ref{nzwide} that in the 18$\umu$m band there are almost no normal galaxies beyond a redshift of 1, thus further simplifying the colour-colour diagram.

\section{Summary and Conclusions}\label{sec:conclusions}

We have introduced ASTRO-F (the Imaging Infra Red Surveyor - IRIS) as a next generation infra-red telescope in space. ASTRO-F has both the capability to take extremely deep pointed observations of 100sq.arcmin. regions of the sky in 9 separate wavelength bands from 2.2-25$\umu$m at mid and near infrared wavelengths down to sensitivities of an order of magnitude below all but the deepest ISO surveys (see tab.\ref{surveys}), at twice the resolution ~\cite{wat00}. 

Two basic survey strategies have been presented. The first, an ultra-deep survey covering $\sim 3200$sq.arcmin. in a doughnut around the North Ecliptic Pole and the second a wide area shallow survey covering almost 18sq.deg..

Survey predictions have been made assuming 2 starkly different evolutionary models including normal, starburst, ultraluminous and Seyfert galaxy components. The first is a classic $(1+z)^{3.1}$ pure luminosity evolution ({\it Levol}) model ~\cite{cpp96}. The second({\it ULIG}) model is a more drastic evolutionary model incorporating a strongly evolving ULIG component ~\cite{cpp002}.

For the deep survey, in general the evolutionary models predict that between 20,000-30,000 sources would be detected in the shortest wavebands dropping to $\approx$5000 in the longest (25$\umu$m) band. 

At the longest MIR wavelengths, the deep survey would be severely constrained by the source confusion limit rendering the deepest integrations relatively futile, although the simultaneous observations with the IRC-NIR will assist in the source extraction . This will be an important constraint for both the ASTRO-F, SIRTF and future relatively small aperture space infrared telescopes. Although SIRTF has a slightly larger primary (85cm) than ASTRO-F, source confusion at longer mid-infrared wavelengths will still be severe. To integrate deeper {\it below} the current constraints set by the source confusion, telescopes with significantly larger mirrors will required ~\cite{cpp003} (e.g. NGST and the proposed HII/L2 mission ~\cite{naka98}).

 In the shorter wavelength bands it should be possible to detect many sources out to high redshift $\sim 5$ with more than half of the normal galaxies being at redshift $>$1 in the 7 \& 9$\umu$m bands. Without doubt the infrared unidentified bands (UIB - PAH features) in the observed galaxy spectra aid in the detection of galaxies to higher redshift, with the 3.3$\umu$m feature enhancing the detectabilty of normal galaxies out to high redshift and the PAH {\it forest} in general enhancing the detectability of all starforming and normal galaxies in the 20$\umu$m band.

The large area, 18sq.deg., shallow survey would produce between 150,000 sources in the shortest waveband to 100,000 sources in the longest waveband. In this case, source confusion is not so much of a problem as only a single pointing is made at each position. Note that a single pointing corresponds to 42-60 individual frames. Therefore there is intrinsic redundancy to distinguish real detections from those of spurious events such as cosmic ray hits which can have a severe effect on IR observations e.g. Aussel et al. ~\shortcite{aussel99}). However, if the stronger evolving {\it ULIG} model is assumed then even a shallow survey will be source confused to some extent at the longest MIR wavelengths (20-25$\umu$m). 

The orbit of ASTRO-F is naturally conducive to the NEP deep survey, allowing many chances of observation every orbit. Where as the shallow area survey would require multiple passes of the same area to allow observations at all wavelengths (i.e. 3 passes). This may incur intolerable overheads. Such overheads could be reduced by using the filters in such a combination as to minimize waste, i.e. 25$\umu$m + 11$\umu$m, 20$\umu$m + 9$\umu$m, 15$\umu$m + 7$\umu$m. In a further response to this problem we have investigated the possibility of using the {\it spare} positions on the IRC-MIR-S and IRC-MIR-L filter wheels as additional {\it wide band\,} filters at 9$\umu$m and 18$\umu$m respectively. Although there is a gain in sensitivity in using these wider band filters, this would be lost on the deep survey as the confusion noise is already above or close to the detection limits. For the shallow survey we find that using this wide-band filter configuration leads to a higher or similar number of detections in the 9$\umu$m and 18$\umu$m respectively.

Using colour-colour plots we should be able to distinguish between different galaxy populations and also obtain rough photometric redhifts if the MIR-S and MIR-L observations are combined with an additional observation using the IRC-NIR channel at 2.2$\umu$m.

During its $\sim 500$ day lifetime (defined by the estimated time of Helium exhaustion), ASTRO-F should make of the order of 7400 pointed observations with the IRC instrument. Of these up to; 32x20 (NEP survey) x 3 (narrow bands filters) + 32x20 (shallow survey) x 2 (wide band filters) = 3200 pointings (or more) could be invested into MIR surveys (since ASTRO-F is in essence a survey mission). Such surveys would approximately double the current area surveyed in the mid-IR (with ISO) and reach to extremely deep fluxes. Following the same doctrine as the ISO 15$\umu$m surveys ~\cite{elbaz98}, ASTRO-F could produce a number of mid-IR surveys covering a wide spectrum of sensitivities and spatial areas. In addition, follow up observations with both the IRC (including spectroscopy) and from the ground, would provide extremely detailed information over the entire MIR wavelength range as to the nature of all types of galaxies in the infrared. Finally it is hoped that ASTRO-F and SIRTF will complement each other and pave the way for future generations of space telescopes in the near \& mid-infrared (NGST, HII/L2).

\section{Acknowledgements}

The ASTRO-F project is managed and operated by the Institute of Space and Astronomical Science (ISAS) Japan in collaboration with the groups in universities and institutes in Japan. We are grateful for all the members of the ASTRO-F project for their efforts and support. The authors would like to thank the anonymous referee for many useful suggestions that greatly improved the clarity of this paper. CPP is supported by a Japan Society for the Promotion of Science (JSPS) fellowship.



\bsp 

\label{lastpage}

\end{document}